\documentclass[english,preprint,showpacs,preprintnumbers,amsmath,amssymb,prb,aps,manuscript]{revtex4-1}
\usepackage[T1]{fontenc}
\usepackage[latin9]{inputenc}
\usepackage{textcomp}
\usepackage{amsmath}
\usepackage{amssymb}
\usepackage[dvipdfm]{graphicx}
\usepackage{natbib}
\usepackage{subfigure}
\usepackage{tabularx}
\usepackage{epsfig}
\usepackage{longtable}
\usepackage{amsfonts}
\usepackage{rotating}

\makeatletter
\newcommand{\be}{\begin{equation}}
\newcommand{\ee}{\end{equation}}
\newcommand{\bea}{\begin{eqnarray}}
\newcommand{\eea}{\end{eqnarray}}
\newcommand{\lyxmathsym}[1]{\ifmmode\begingroup\def\b@ld{bold}
  \text{NHGLE\ifx\math@version\b@ld\bfseries\fi#1}\endgroup\else#1\fi}

\@ifundefined{textcolor}{}
{%
 \definecolor{BLACK}{gray}{0}
 \definecolor{WHITE}{gray}{1}
 \definecolor{RED}{rgb}{1,0,0}
 \definecolor{GREEN}{rgb}{0,1,0}
 \definecolor{BLUE}{rgb}{0,0,1}
 \definecolor{CYAN}{cmyk}{1,0,0,0}
 \definecolor{MAGENTA}{cmyk}{0,1,0,0}
 \definecolor{YELLOW}{cmyk}{0,0,1,0}
 }

\makeatother

\usepackage{babel}

\begin{document}

\title{Simulation of quantum zero point effects in water using a frequency dependent thermostat}

\author{Sriram Ganeshan\textsuperscript{1},  R. Ram\'irez\textsuperscript{2}, M. V. Fern\'andez-Serra$^{1,a}$
 \let\oldthefootnote\thefootnote\global\long\def\thefootnote{{a)}}
\footnotetext{Electronic mail: maria.fernandez-serra@stonybrook.edu}\let\thefootnote\oldthefootnote}

\affiliation{\textsuperscript{1}Department of Physics and Astronomy, Stony
Brook University, Stony Brook, New York 11794-3800, USA}

\affiliation{\textsuperscript{2}Instituto de Ciencia de Materiales de Madrid
(ICMM), Consejo Superior de Investigaciones Cient\'ificas (CSIC), Campus
de Cantoblanco, 28049 Madrid, Spain }

\date{{\today}}

\begin{abstract}

Molecules like water have vibrational modes with zero point energy well above room temperature.
As a consequence, classical molecular dynamics simulations of their liquids largely underestimate the energy of modes with higher zero-point temperature, which translates into an underestimation of covalent interatomic distances due to anharmonic effects.
Zero point effects can be recovered using path integral molecular dynamics simulations, but these are computationally expensive,
making their combination with ab-initio molecular dynamics simulations a challenge.
As an alternative to path integral methods, from a computationally simple perspective, one would envision the design of a thermostat capable of
equilibrating and maintaining the different vibrational modes at their corresponding
zero point temperatures. Recently, Ceriotti {\it et al.} [Phys. Rev. Lett. 102, 020601 (2009)] introduced a framework to use a custom-tailored Langevin equation with correlated-noise that can be used to include quantum fluctuations in classical molecular-dynamics simulations.
 Here we show that it is possible to use the generalized Langevin equation with suppressed noise in combination with Nose-Hoover thermostats
 to efficiently impose zero-point temperature to independent modes in liquid water. Using our simple and inexpensive method, we achieve excellent agreement on all atomic pair correlation functions compared to the path integral molecular dynamics simulation. \end{abstract}
\maketitle

\section{Introduction}

Understanding how large are zero point nuclear quantum effects (NQE) both in water\cite{Manolopoulos09,vega10, paesani10, briesta10, zheng11,Ramirez2010a,Car08,Ramirez11} and ice\cite{Pamuk2012,Galli2012,Lin2011} is an active area of research. Recently, the anomalous isotope effect on the thermal expansion of ice\cite{Pamuk2012,Kuhs94}, was explained using first
principle simulations. The effect is anomalous because of two different properties. The first anomaly shows that the crystal with lighter isotopes (hydrogen) has smaller volume than
the crystal with heavier isotopes (deuterium). The second, and even more strange one, shows that the volume difference between H$_2$O and D$_2$O increases with temperature. 
The two anomalies were shown to be originated on competing anharmonicities in the vibrational modes of the system. 
However, the question of how much this effect will play a role in the liquid phase is still open.
Experimentally, there are indications\cite{Mayers12} that quantum effects might be different in low density and high density regions of liquid water.
Understanding to which point this is correct will shed light on the question
of how much the local environment of the proton will modify its kinetic energy.

At the theoretical level, there are still open questions regarding how much the structure of liquid water is dependent on the classical treatment of the ionic degrees of freedom in {\it ab-initio} molecular dynamics (MD) simulations is an open question\cite{Car08,car10,Lin2011,Kong12, Pettersson10}. Even if a number of path integral molecular
dynamics studies have directly addressed the issue\cite{Car08,car10,Manolopoulos09,Ramirez2010a,Ramirez11,Lin2011}, a definite answer has not yet been provided.
The problem is subtle, due to the complex nature\cite{mvfsprl06} of the OH--O hydrogen bond (Hbond) in water.
It is well known that hydrogen bonded materials show an anti-correlation\cite{Libowitzky99} between the high energy, stretching frequencies and
the librational frequencies of the molecules. Recently\cite{Pamuk2012}, we have shown that this anti-correlation is the origin of negative gr\"{u}neisen parameters of the high energy vibrational modes in ice. These are large enough to cause an anomalous isotope effect in the volume of ice, making the volume per molecule of heavy or D$_2$O ice larger than that of normal or H$_2$O ice.
This anomaly is not captured by flexible and/or polarizable force-fields, due to their underestimation of the anti-correlation effect\cite{Pamuk2012}.
Nonetheless, we choose to use in this study the q-TIP4P/F\cite{Manolopoulos09} force field. Even if it has been shown to fail in the description of
the anomalous isotope effect of ice\cite{Pamuk2012}, it provides a good qualitative description of the anharmonicities of all the modes in liquid water.
In classical MD simulations of force field models, all the modes are equilibrated at a given constant temperature. This equipartition of temperature is a classical description of liquid water which lacks NQE. Recently, Ceriotti {\it et al.}\cite{ceriotti1,ceriotti2,ceriotti3,ceriotti4,ceriotti5,ceriotti6} have shown that the key features of path integral molecular dynamics (PIMD) simulations of liquid water can be reproduced using custom tailored thermostats based on generalized Langevin dynamics (GLE). In their work, they were able to enforce the $\omega$-dependent effective temperature $T(\omega)=\frac{\hbar \omega}{2k_B}\coth\frac{\hbar \omega}{2k_B T} $ simultaneously on different normal modes, without any explicit knowledge of the vibrational spectrum.  The tailoring aspect of their thermostat involves complicated optimization to independently tune the drift and diffusion parameters of the GLE dynamics.
In this work we propose to analyze the problem by introducing a thermostating scheme with very few, and easy to tune parameters that can equilibrate modes to different temperatures. In our scheme we couple both Nose-Hoover (NH) and GLE dynamics to the system. We use GLE kernels that satisfy the fluctuation-dissipation (FD) condition which can be derived from a well defined harmonic bath model. We suppress the noise term in GLE dynamics by setting the GLE temperature to $0$. In this limit, the dynamics is deterministic. The frequency dependent equilibration is achieved through the independent tuning of NH and the frequency dependent friction profile. The NH thermostat brings out the role of the non-local friction profile while the noise term remains suppressed. Microscopic details of the full dynamics are presented in Appendix.~(\ref{sec:glemodel}).
We sacrifice transferability of parameters between different systems in exchange for simplicity in their optimization against the known vibrational spectrum of the system.
  This thermostat acts on the system within a deterministic regime and hence our method can be thought as a deterministic frequency dependent thermostat or phonostat\cite{Phonostat-Grossman}.
The goals of this study are two sided. On the methodological side, after rigorously deriving the thermostat equations, we evaluate its performance , by comparing it with
 PIMD simulations of q-TIP4P/F water.
In addition, we address the question of competing quantum effects\cite{Manolopoulos09} or competing anharmonicities in water\cite{Pamuk2012}
using a quantified, temperature-dependent approach. To achieve this we reformulate the idea of NQE in terms of the zero point energy of individual modes.

\section{Simulation Details}

In this work, we construct a frequency dependent thermostating scheme. This scheme involves use of two thermostats. We use the standard NH chain thermostats which is the gold standard of thermostats. To achieve frequency dependent equilibration, we use GLE to modify the NH action. Recently, Ceriotti and Parinello\cite{ceriotti1,ceriotti2,ceriotti3,ceriotti4,ceriotti5,ceriotti6} developed an extensive thermostating scheme based on GLE dynamics. We choose a particular form of GLE dynamics and use it in conjunction with the NH dynamics to enforce frequency dependent thermostating. We call this scheme NHGLE (for Nose-Hoover-GLE) thermostating. 
The microscopic picture of this scheme is outlined in the appendix.

\begin{center}
\begin{figure}[htb!]
\includegraphics[width=8.2cm,height=5cm]{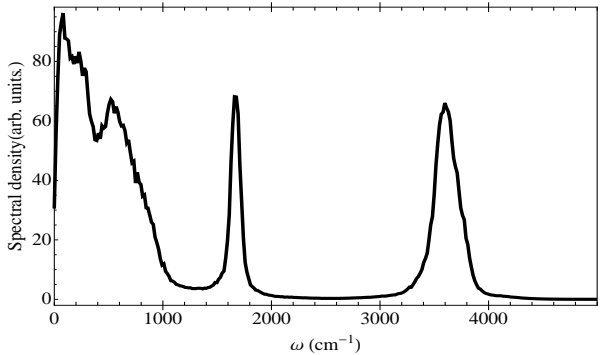}
\caption{Spectral density plot of q-TIP4P/F liquid water equilibrated with NH thermostats at $300$ K.  }
\label{spectrumnvt}
\end{figure}
\end{center}

\begin{center}
\begin{figure}[ht]
\includegraphics[width=8.2cm,height=5cm]{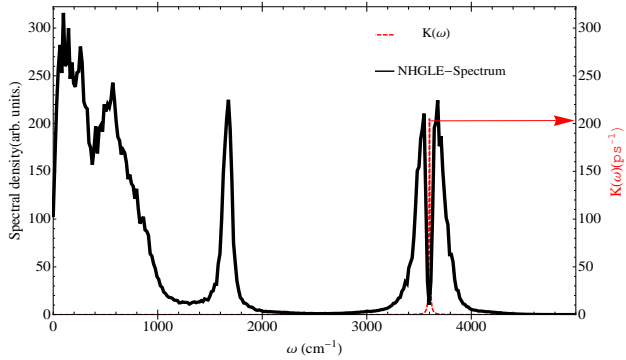}
\caption{(Color online) Black line: Spectral density plot of q-TIP4P/F liquid water at $300$ K with NHGLE. Red line: delta-Like peak
frequency dependent friction profile. The axes ticks on the left edge corresponds to the spectral density. The axes ticks on the right edge of the frame corresponds to $K(\omega)$ in units of $ps^{-1}$}.
\label{narrowmodes}
\end{figure}
\end{center}

The implementation of NHGLE thermostat is as follows. We begin by writing the classical evolution for the full dynamics derived in Appendix(\ref{sec:glemodel}). The infinitesimal evolution operator is given as,
\bea
U(\Delta
t)&=&e^{i\Delta t(\textsl{L}_{system}+\textsl{L}_{NH}+\textsl{L}_{GLE})}\\
&\approx &e^{i\frac{\Delta t}{2}(\textsl{L}_{GLE}+\textsl{L}_{NH})}e^{i\Delta t\textsl{L}_{system}}e^{i\frac{\Delta t}{2}(\textsl{L}_{GLE}+\textsl{L}_{NH})}\nonumber\\
\eea
Where $\textsl{L}$ is the Liouville operator. The integrators for MD can be obtained from the Trotter factorization of Liouville propagators. The corresponding Liouville operators for NH and GLE are given in detail in Refs.~(\onlinecite{tuckerman1990,ceriotti3}) respectively. The evolution of NH and GLE is updated at $\Delta t/2$ before and after the velocity-verlet routine for system evolution. As described in Appendix~\ref{sec:glemodel} we have two thermostats acting on the system  and they compete to enforce their respective temperatures. The strength of the GLE thermostat varies with the frequency and is strongest for the modes in the range $ \approx \omega_{0} \pm \Delta \omega $. The result of this competition is that the modes in the range $ \approx \omega_{0} \pm \Delta \omega $ are thermostated at an effective temperature $T_{eff}<T_{NH}$. Depending on the strength of the friction coefficient (height of $K(\omega)$ peak), the effective temperature of the specific modes can lie anywhere between  $0 \le T(\omega)_{eff} \le T_{NH}$. Temperature of all the other modes are equilibrated at $T \sim T_{NH}$ as the $K(\omega)\sim0$ for $ \omega \notin (\omega_{0}-\Delta \omega,\ \omega_{0}+\Delta \omega $). Hence we can control the effective temperature of a particular normal mode using NHGLE dynamics.
Our system consists of 256 water molecules with a density of 0.997 $g/cm^3$ modeled by the flexible force field model q-TIP4P/F\cite{Manolopoulos09}. All simulations have been performed at the time step of $0.5fs$.
 This model has been used in many isotope effects studies both for water and ice\cite{Ramirez2010a,Ramirez11,Ramirez12}, and therefore is an excellent model to evaluate the performance of our approach.
Before imposing the complete zero-point temperature distribution in water, we demonstrate effects of damping of narrow range of modes using NHGLE thermostats. This example will establish the spirit in which we intend to use NHGLE thermostats. We use in house developed code for the force field and MD implementations. In Fig.~\ref{spectrumnvt} we show the vibrational density of states (obtained from the Fourier transform of the velocity autocorrelation function) for the q-TIP4P/F model. The three peaks corresponds to translation+rotations ($400-1000~cm^{-1}$), bending ($\sim1600~cm^{-1}$) and stretching ($\sim3600~cm^{-1}$) modes. We also plot the projected temperature of translation, rotation and intramolecular vibrational modes(see Fig. (\ref{tempall}(a))). Mode-projected temperatures is an important parameter to monitor NHGLE action on the system.
 We calculate these projections by defining new molecular subspaces along the center of mass (for the translations), molecular main three moment of inertia axis (for the rotations) and the three vibrational normal modes of the isolated molecule. These projections act as a guide to tune NHGLE parameters for designing a frequency dependent temperature control.
For GLE implementation we base our code on codes developed by Ceriotti {\it et al.}\cite{note2}.   We see that without the GLE action all the modes are equilibrated to the temperature set by NH thermostats. We code the memory kernels to overlap with the vibrational spectrum of the system and tailor them to our requirement. We select a memory profile of very narrow width sharply peaked at some frequency, $\omega_0$ which vanishes rapidly away from it. NH chains are set to $300$ K. GLE is much more strongly coupled to the system than the NH chains at selective modes ($\omega_0 \pm \Delta \omega$). In the expression of memory kernel, Eq.~(\ref{kernel}), we set the peak position $\omega_0=3600~cm^{-1}$, peak width  $\Delta\omega=5~cm^{-1}$ and the strength of coupling $\gamma^{-1}=1000$ ps (see Fig. \ref{narrowmodes}).
 We now extend this method to study how the structure of liquid water changes when modes are kept at different temperatures.
\section{NHGLE applied to q-TIP4P/F water}\label{sec:weakrdf}
Using NHGLE to simulate liquid water to non-equilibrium distribution of temperatures, we can analyze how the structure of q-TIP4P/F water depends on the temperature of individual modes. This way we can establish the influence of each individual mode in the structure of the water. Also one can study how the different dynamical modes are connected to each other.
\subsection{Weak damping of intramolecular modes}
\begin{center}
\begin{figure}[htb!]
\includegraphics[width=8.2cm,height=5cm]{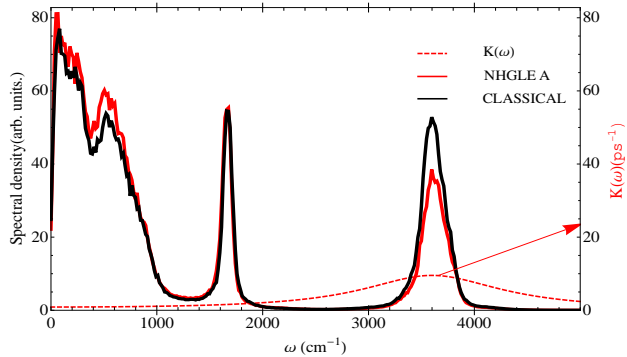}
\caption{(Color online) Spectral density plot. The black line is for of q-TIP4P/F liquid water at $300$ K, the solid red line
is for the NHGLE A (see  Table~\ref{tab:pterms}). The red dashed line shows the frequency dependent friction profile. The axes ticks on the left edge corresponds to the spectral density. The axes ticks on the right edge of the frame corresponds to $K(\omega)$ in units of $ps^{-1}$.}
\label{specrum_vibdamp}
\end{figure}
\end{center}

We damp the intramolecular modes and study their influence on the overall structure of liquid water,  using a damping profile as shown in Fig. (\ref{specrum_vibdamp}). For this simulation (NHGLE A) parameters are described in Table~\ref{tab:pterms}. Projected temperatures plot (Fig~\ref{tempall}(b)) show that the intramolecular modes are kept are relatively lower temperatures for the full simulation run.
 We also plot the radial distribution function (rdf) for NHGLE A. This rdf  clearly shows a local softening of the first two O-O peaks. On the other hand the first peak of the  O-H rdf as expected is much sharper. The higher order peaks, linked to the Hbond network have also become softer.  A more frozen covalent bond results in a loss of structure of liquid water.
This is a consequence of the previously mentioned anticorrelation. The Hbond is weakened by strengthening the O-H covalent bond.

\begin{center}
\begin{figure}[htb!]
\includegraphics[width=8.2cm,height=5cm]{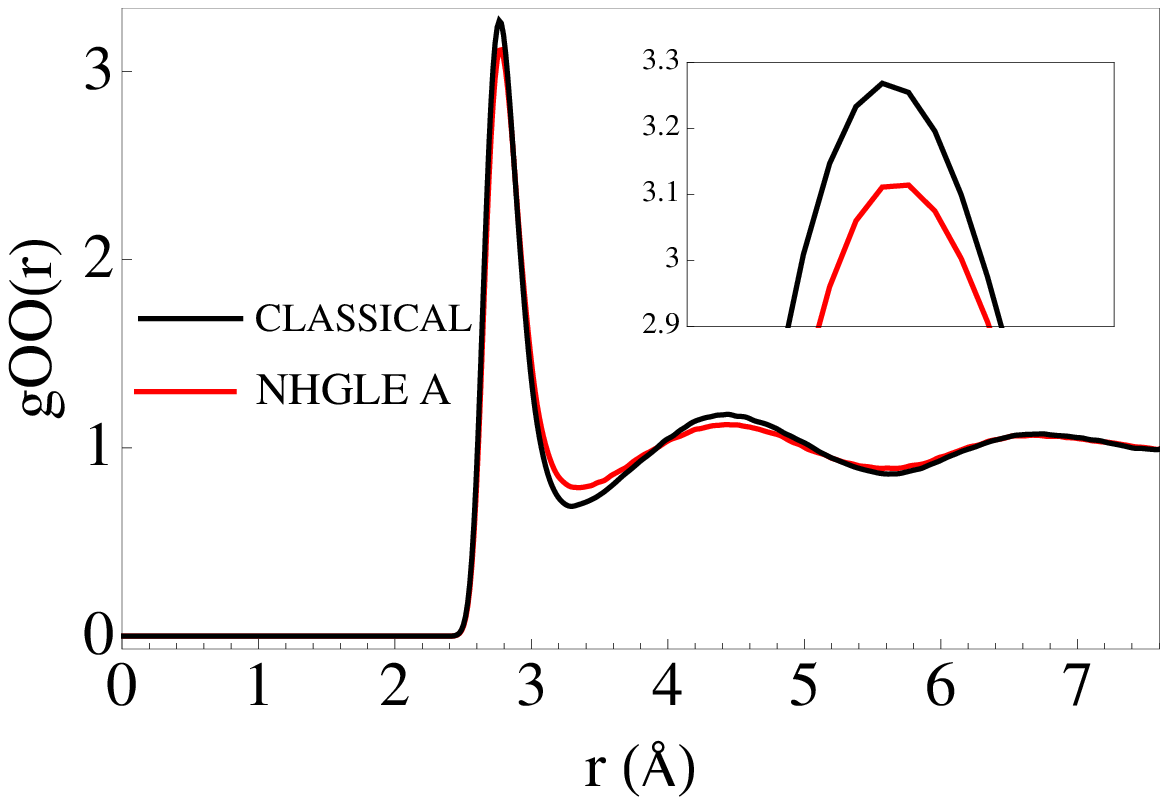}\\
\vspace{1cm}
\includegraphics[width=8.2cm,height=5cm]{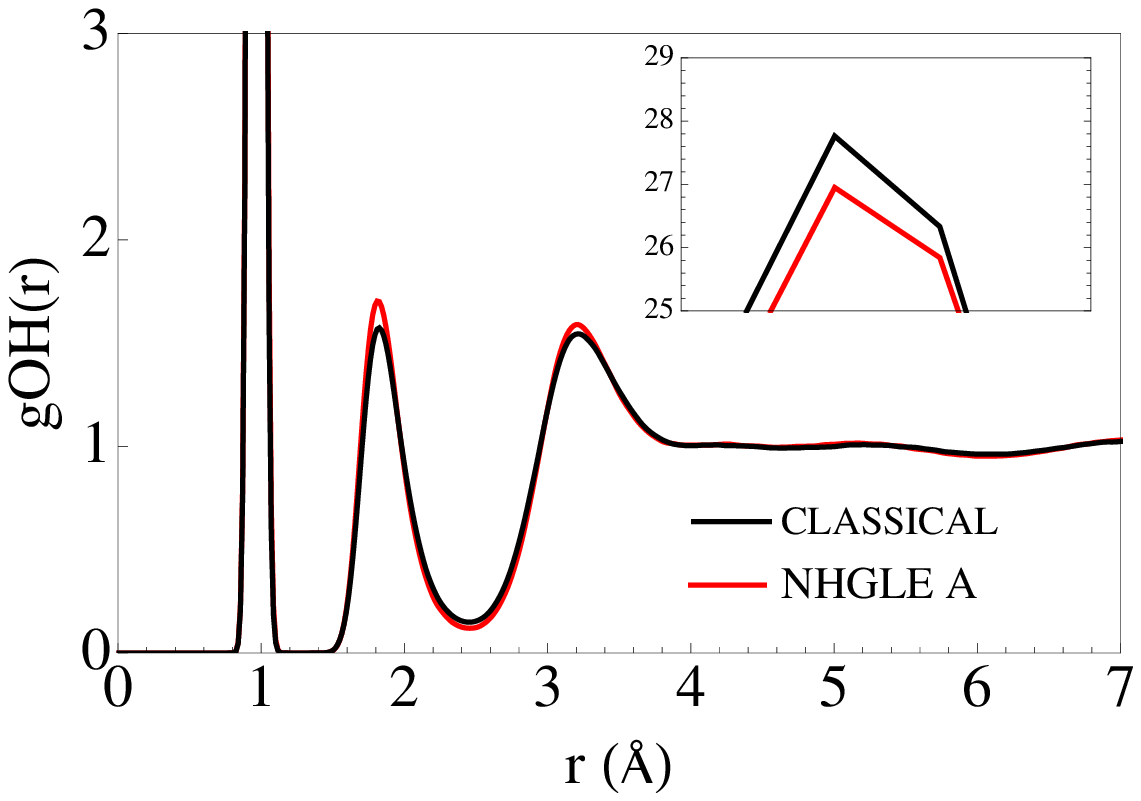}
\caption{(Color online)Radial distribution function plots of q-TIP4P/F liquid water for NVT simulation (black) at $300$ K and NHGLE damped OH stretching (red).
 Top: O-O rdf. Bottom: O-H rdfs, the inset shows a zoom into the first peak. }
\label{dampoh}
\end{figure}
\end{center}

\subsection{Weak damping of intermolecular modes}
\begin{center}
\begin{figure}[htb!]
\includegraphics[width=8.2cm,height=5cm]{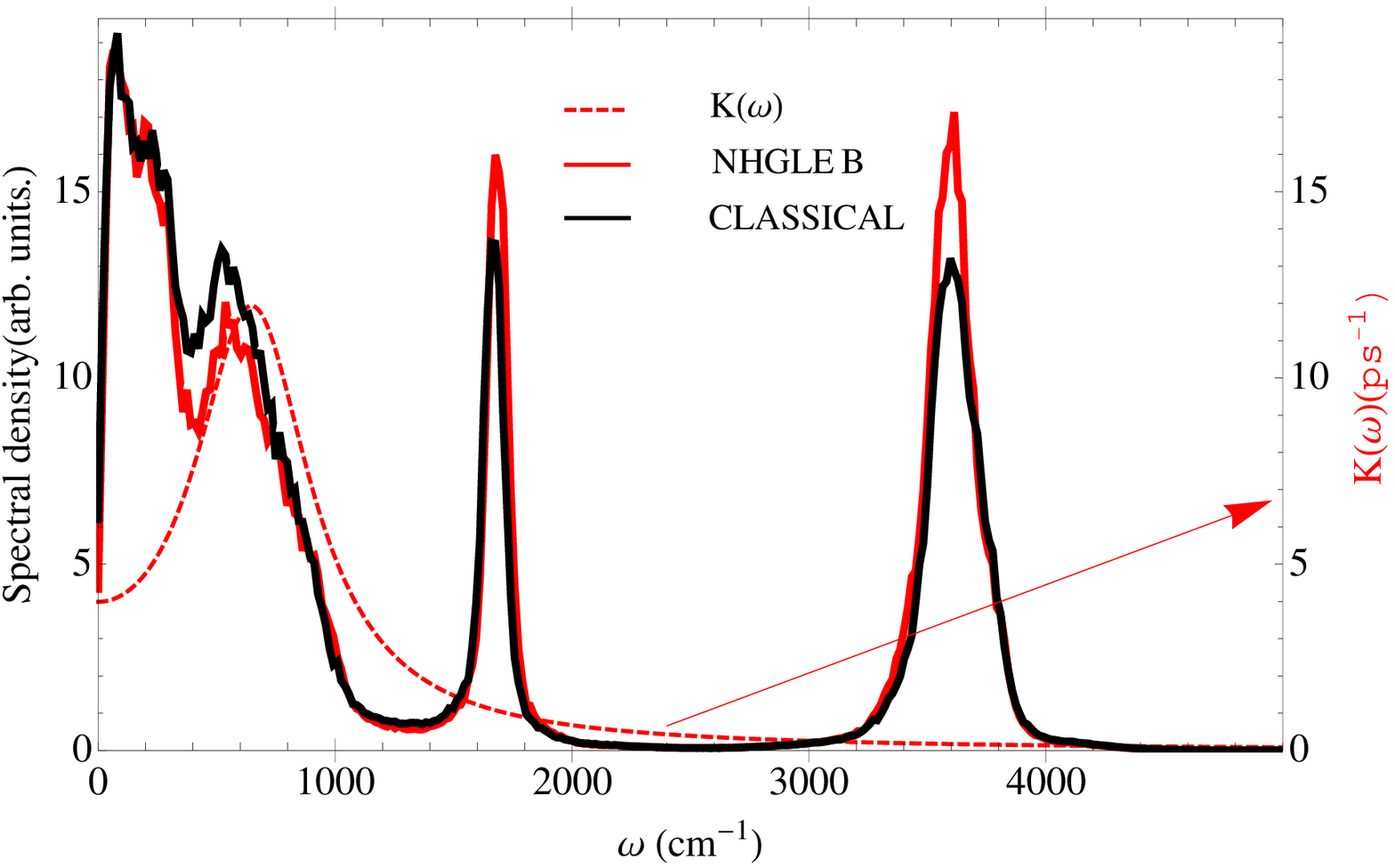}
\caption{(Color online) Spectral density plot. The black line is for of q-TIP4P/F liquid water at $300$ K, the solid red line
is for the NHGLE B (see  Table~\ref{tab:pterms}). The red dashed line shows the frequency dependent friction profile. The axes ticks on the left edge corresponds to the spectral density. The axes ticks on the right edge of the frame corresponds to $K(\omega)$ in units of $ps^{-1}$.}
\label{specrum_libdamp}
\end{figure}
\end{center}

\begin{center}
\begin{figure}[htb!]
\includegraphics[width=8.2cm,height=5cm]{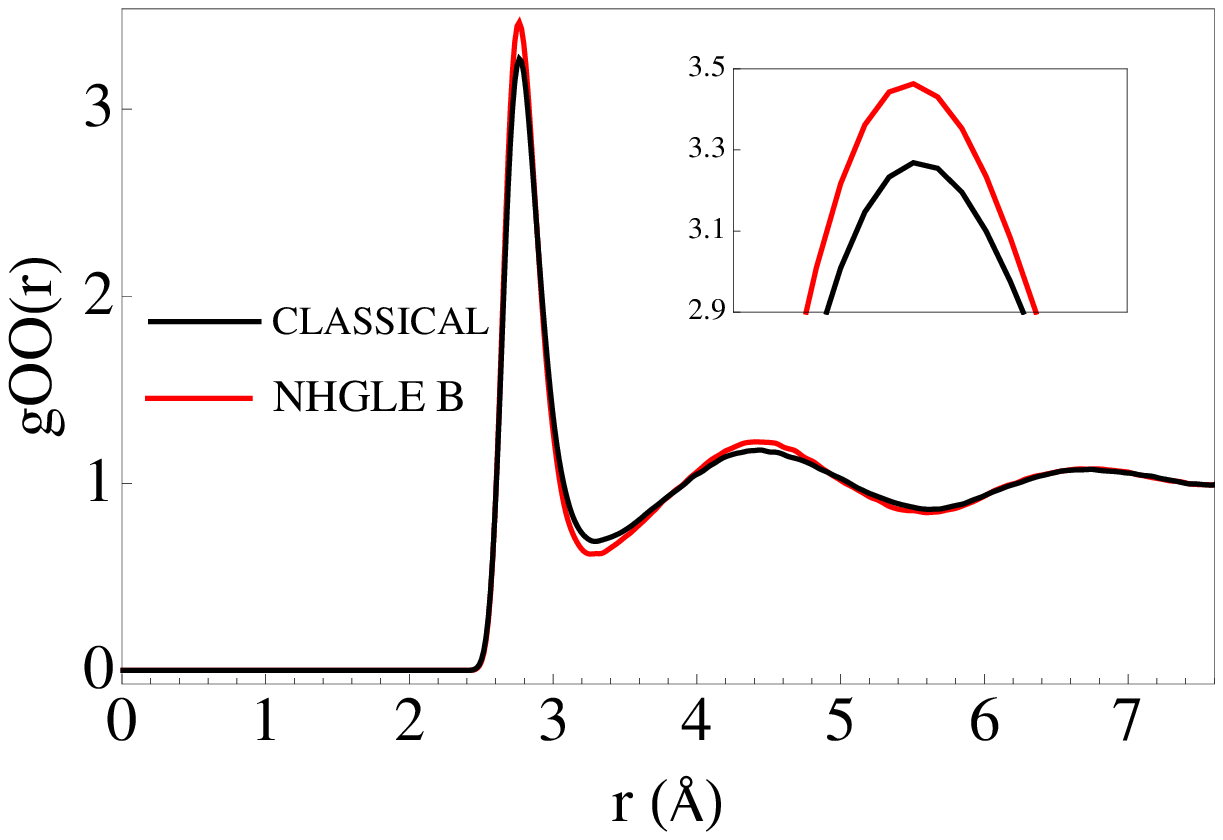}\\
\vspace{1cm}
\includegraphics[width=8.2cm,height=5cm]{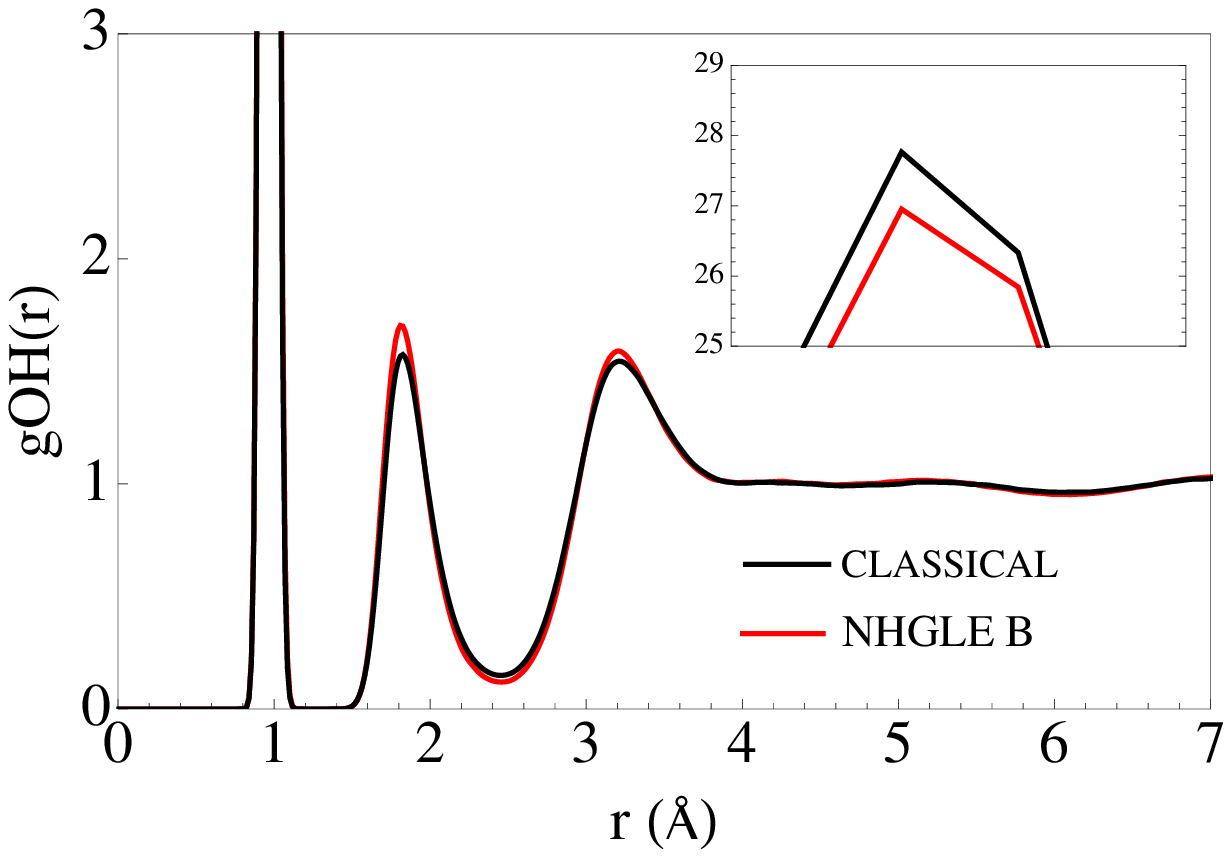}
\caption{(Color online) Radial distribution function plots of q-TIP4P/F liquid water for NVT simulation (black) at $300$ K and NHGLE damped
intermolecular stretching (red).
 Upper panel: O-O rdf. Lower panel : O-H rdfs, the inset shows a zoom into the first peak. }
\label{dampnet}
\end{figure}
\end{center}

\begin{center}
\begin{figure}[htb!]
\includegraphics[width=9.2cm,height=5.5cm]{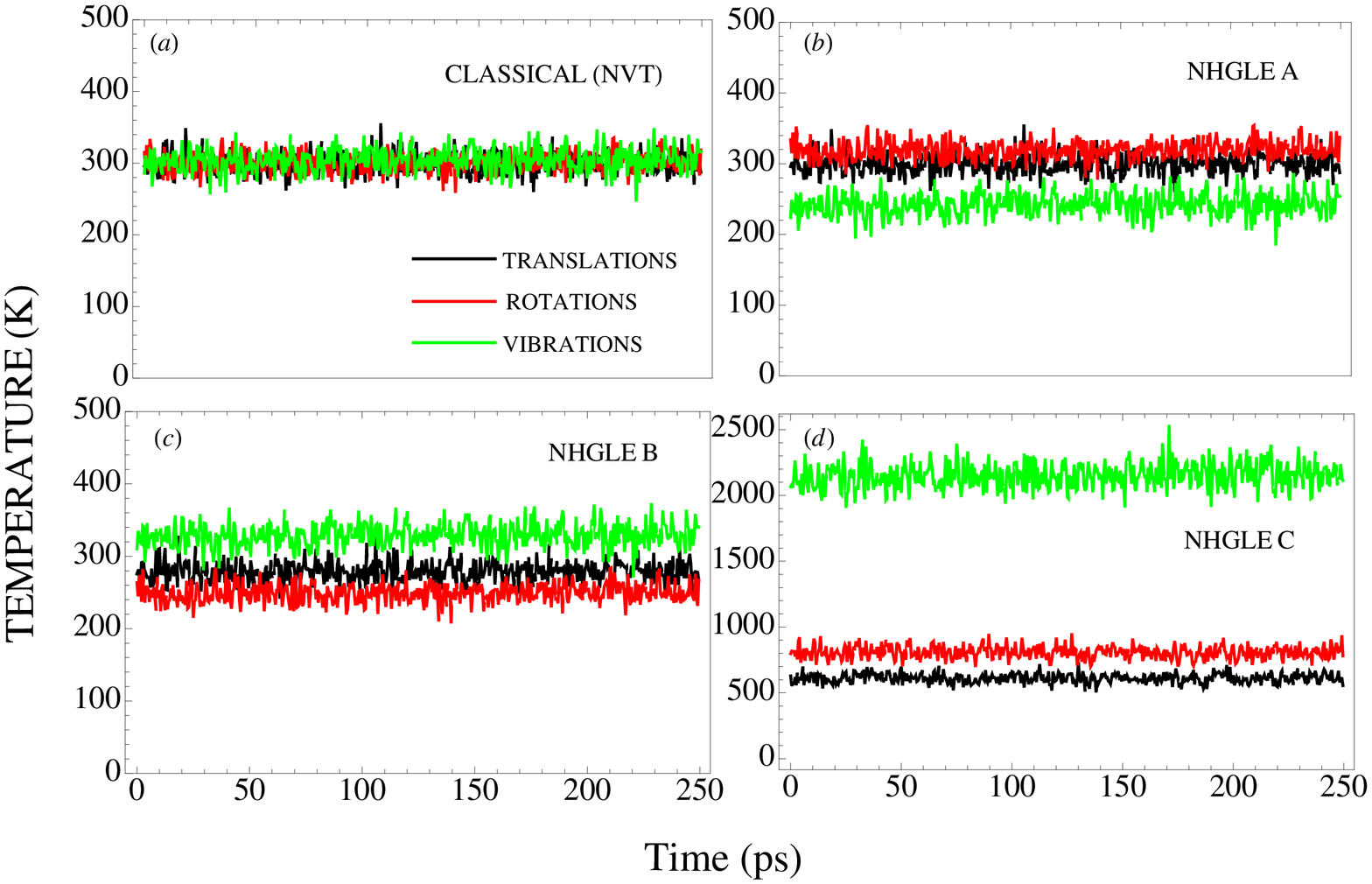}
\caption{(Color online)Temperature of translational (black), rotational (red) and vibrational (green) modes in liquid water plotted as a function of time. This mode projected temperature is plotted for (a) NVT simulation at 300 K. (b) NHGLE A simulation with damped intramolecular OH stretching . (c) NHGLE B simulation with damped intermolecular modes. (d) NHGLE C simulation with modes close to effective zero point temperature. See  Table (\ref{tab:pterms}) for temperature values.}
\label{tempall}
\end{figure}
\end{center}

In this section we analyze whether a more fluctuating O-H covalent bond induces a local structuring of liquid water. To answer this question we now damp the low energy intermolecular modes.  We couple GLE to low energy modes with almost no coupling to intramolecular modes.
The parameters for this simulation (NHGLE B) are described in Table~\ref{tab:pterms}
The Results are shown in Figs.~(\ref{specrum_libdamp},\ref{tempall}).
The action of damping low energy modes results in relatively higher temperature of vibrational modes (see Fig.~\ref{tempall}(c)). Consequently, we see more structure in the O-O rdf (see Fig. \ref{dampnet}). This establishes the well know anticorrelation\cite{Pamuk2012} between inter and intramolecular modes in liquid water. Hence we demonstrate that the strengthening of the water structure (by more frozen Hbonds) implies a more delocalized O-H intramolecular bond, i.e. a softening of the stretching OH vibration. In Table (\ref{tab:pterms}) we present normal mode-projected temperatures for translations, rotations and vibration modes.

\
\begin{table*}[!hbtp]	
\centering
\caption{Temperature distribution among individual modes}
\resizebox{12cm}{!}{
\begin{tabular}{ccccccccc}
	\hline
	   & \multicolumn{1}{c|}{$T_{NH}$(K) } &  \multicolumn{1}{c|}{Trans(K) }&  \multicolumn{1}{c|}{Rot(K) }&  \multicolumn{1}{c|}{Vib(K) } &  \multicolumn{1}{c|}{ $\omega_0$ (cm$^{-1}$) } & \multicolumn{1}{c|}{ $\Delta \omega$ (cm$^{-1}$) } & \multicolumn{1}{c|}{ $\gamma^{-1}$ (ps) }  & \multicolumn{1}{c|}{ $T_{GLE}$(K) }     \\
		\hline \hline
    NVT     &300  & 299  & 300 & 302 & -  & - & - & -\\
    NHGLE A  &300  & 295  & 325 & 240 & 3600  & 800 &  0.9 &  0\\
    NHGLE B  &300  & 295 &  270 & 325  & 700  & 300 & 0.2 &  0\\
    NHGLE C  &1600  & 600 & 800  & 2200 & 700  & 650 & 0.006 & 0 \\
\hline\hline
\end{tabular}
}
\label{tab:pterms}
\end{table*}

\section{Quantum simulation with modes close to their zero point temperature}\label{sec:zeropoint_gle}

None of the results shown in the previous sections were surprising, they are a confirmation of the anti-correlation effect. This effect is a manifestation of
the strong anharmonicity of the vibrational modes, which strongly couple to the rotational modes when Hbonds are formed. However, when all the
normal modes are equilibrated at their corresponding zero point temperatures, the two opposite effects
we just described should balance out. As shown by Habershon {\it et al.}\cite{Manolopoulos09}, this competition results
in an overall minimization of quantum effects on the structure of liquid water. To evaluate this using our method, we set the NH-temperature close to the zero point temperature of vibrational modes and damp intermolecular modes so that the effective temperature of the low energy modes are close to their effective zero point temperature. The shape of the tailored frequency dependent memory profile is shown in Fig. (\ref{vvzp}).  The advantage we have is that very few parameters are used to achieve this non equilibrium $T$ distribution.

\subsection{Strategy to design a quantum thermostat using NHGLE}

\begin{center}
\begin{figure}[htb!]
\includegraphics[width=9.1cm,height=6.0cm]{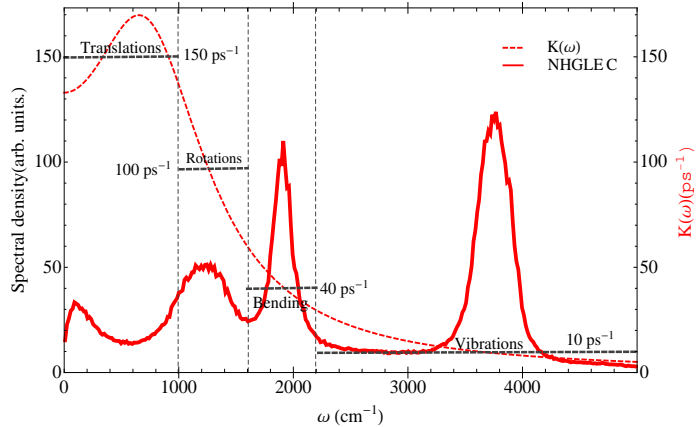}
\caption{(Color online) Spectral density plot, decomposed into different regions of dynamical modes. The horizontal dashed line indicates the average value of friction in each section. The dashed red line interpolates between the dashed lines across different modes. The red line is the NHGLE affected spectrum. The axes ticks on the left edge corresponds to the spectral density. The axes ticks on the right edge of the frame corresponds to $K(\omega)$ in units of $ps^{-1}$.}
\label{vv_breakup}
\end{figure}
\end{center}

In this section we outline the strategy to tune NHGLE parameters to equilibrate modes to their zero point temperature. We emphasize that the fundamental idea is to equilibrate modes to their zero-point temperatures. Water has three distinct vibrational spectrum regions associated to the translational, rotational and vibrational (bending+stretching) motions.
The procedure to build a quantum thermostat from NHGLE- involves the following steps:

\begin{itemize}
\item {\bf Step 1}: knowledge of the vibrational spectrum of the force field used in the simulation is required. The classical spectrum of TIP4P/F shown in Fig.~(\ref{spectrumnvt}) has  three features corresponding to translation+rotations ($400-1000~cm^{-1}$), bending ($\sim1600~cm^{-1}$) and stretching ($\sim3600~cm^{-1}$) modes. We estimate the average zero point temperature ($T_{zp}$) corresponding to these modes using the relation $T_{zp}=\frac{\hbar \omega}{2 k_b}$. The estimated zero point temperature is given as: translations~320K, rotations~570K and vibrations~2600K. The zero point temperature of low energy modes may increase due to the shift in intermolecular modes depending on the degree of anharmonic coupling between the modes. This can be accounted for (as shown in the case of water in this work) by setting an effectively higher temperature on the intermolecular modes in accordance with the modified spectrum (see Fig.~(\ref{vv_breakup}) and Fig.~(\ref{vvzp}) ).
\item{\bf Step 2}: The second step involves fixing the starting parameters ($\gamma, \Delta \omega, \omega_{0}$ and $T_{NH}$) to equilibrate modes to their respective zero-point temperatures. 
\begin{itemize}
 \item{$T_{NH}$:}  We set the NH temperature ($T_{NH}$) close to the zero point temperature of the mode with highest zero-point temperature (in this case vibrations). Since the NH thermostat will equilibrate all modes to $T_{NH}$ we need to use the other three parameters of the friction profile ($\gamma, \Delta \omega, \omega_{0}$ appearing in memory kernel $K(\omega)$ as shown in Eq.~(\ref{kernel})) to damp the low energy modes selectively. We generally set the value of $T_{NH}$ few hundred kelvins less than the zero point estimate of the highest zero point temperature of the system. This is done to account for the energy transferred from the damping of the low energy modes which adds to the temperature of the undamped modes. The results are not very sensitive to small changes (of the order of 100K-200K) to the zero point temperature of the vibrational modes. 
 \item{$\omega_0$:} The peak position of the kernel $K(\omega)$ is at $\omega_0$ (point of highest friction). This the simplest kernel parameter to fix. It is coincided with the 
highest frequency of  the translational modes region of the spectrum. All these modes need to be kept at the lowest zero-point temperature. As the friction decreases steadily  for $\omega>\omega_0$ the rotational modes are automatically kept at a relatively higher temperature than the translational modes.
 \item{$\Delta\omega$:} This parameter controls the width of the friction profile. The translational modes extend all the way to very low frequencies and it is important to keep them damped. Setting $\Delta\omega \lesssim \omega_0$ ensures that the full range of translational modes are damped with constant friction.  $\Delta \omega$ also determines how steeply the friction value goes to zero for $\omega>\omega_0$.
  \item{$\gamma$:} $\gamma$ determines the strength of the friction or the height of the profile $K(\omega)$. It is also determined by the value of $T_{NH}$. The higher the value of  $T_{NH}$, the higher the value of $\gamma$ should be. Typically for $T_{NH}=1600K$, $\gamma\sim170ps^{-1}$ which is also the peak friction at $\omega_0$. 
   \end{itemize}
   \item{\bf Step 3}: All the parameters are fine tuned by  monitoring the mode decomposed temperatures (see Fig.~(\ref{tempall})). The power of this method is that even a very short time simulation of 500fs (250fs equilibration and 250fs run) can accurately reflect the temperature distribution for any length of simulation.  
Therefore the fitting of the friction profile does not require extensive molecular dynamic simulations, but simulations of the order of magnitude of those
used to pre-equilibrate the system using standard thermostats.
    \end{itemize}
    
  \section{Radial distribution functions}
\begin{center}
\begin{figure}[htb!]
\includegraphics[width=8.2cm,height=5cm]{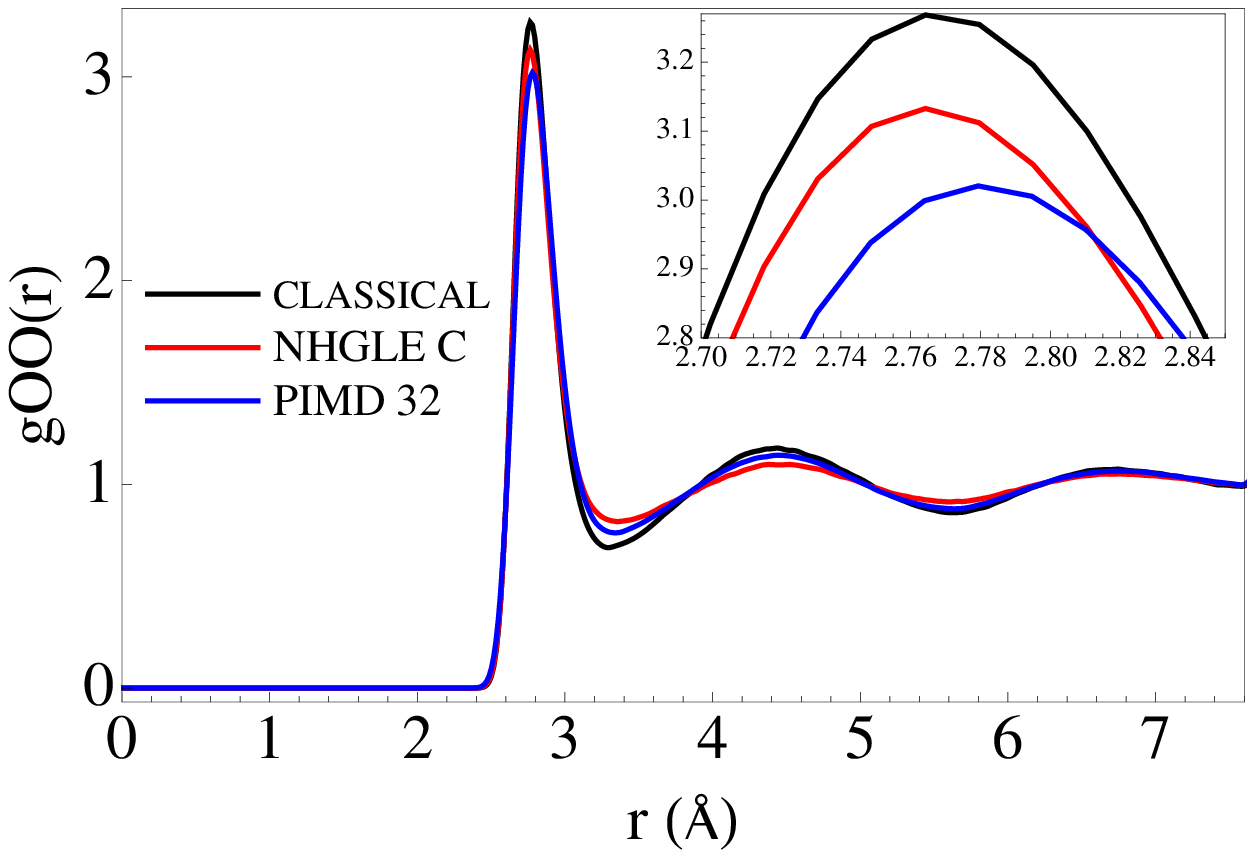}\\
\vspace{0.2cm}
\includegraphics[width=8.2cm,height=5cm]{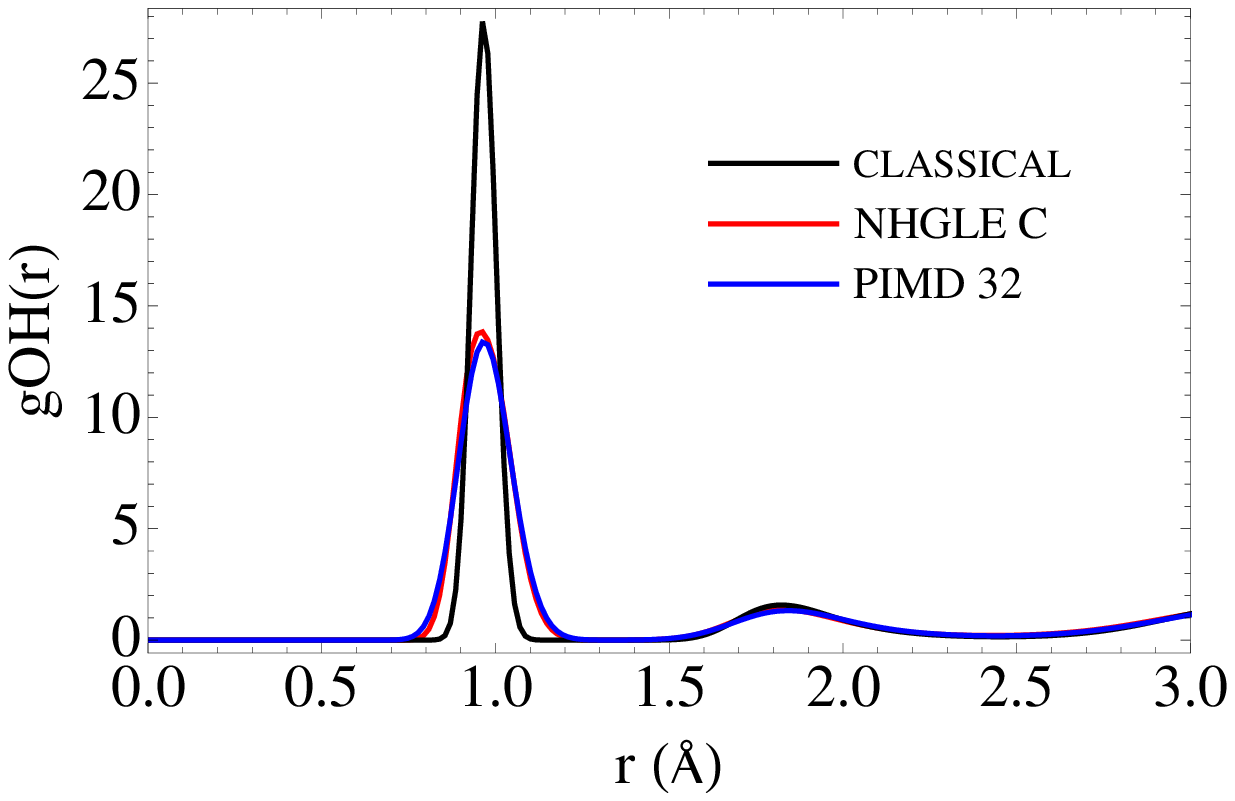}\\
\vspace{0.2cm}
\includegraphics[width=8.2cm,height=5cm]{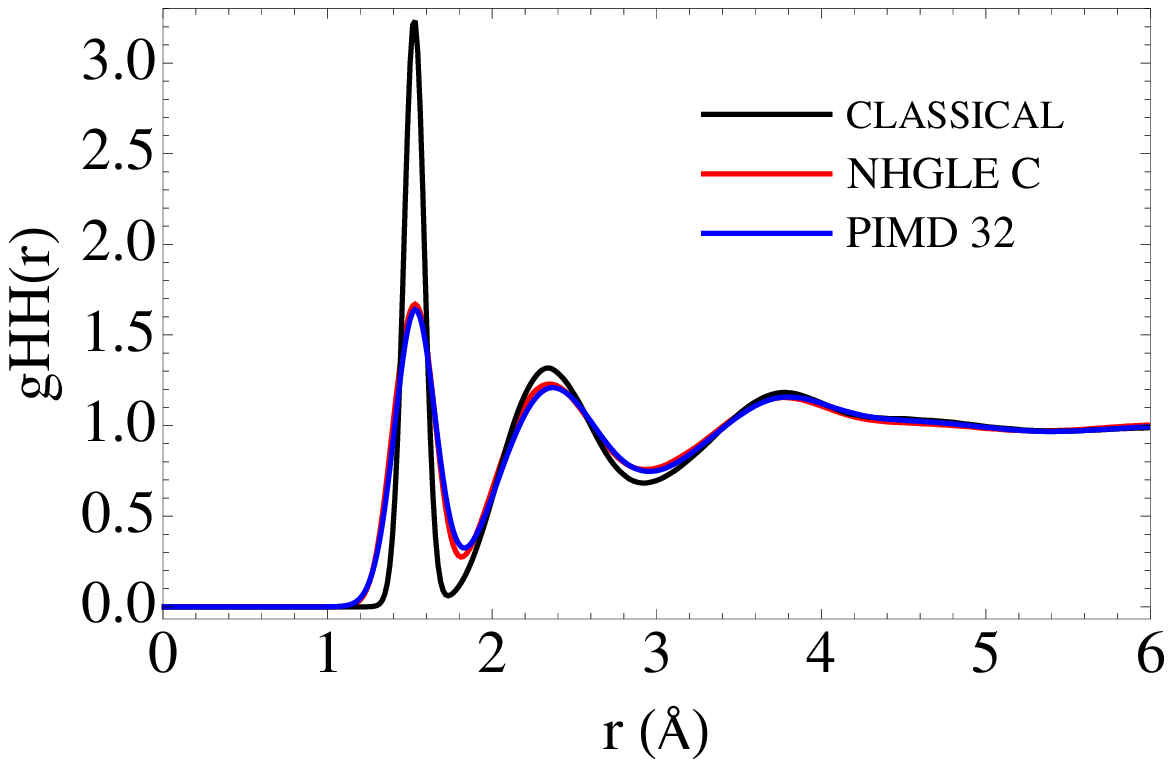}\\
\caption{(Color online)Radial distribution plots. Top: O-O rdf. Middle O-H rdf. Bottom H-H rdf.
 Black line: NVT simulation at 300 K. Red line: Zero-point NHGLE simulation (NHGLE C, in Table~(\ref{tab:pterms})).
Blue line: PIMD simulation with 32 beads.}
\label{rdfzp}
\end{figure}
\end{center}

 We plot the  radial distribution functions of liquid water for this case (see Fig. (\ref{rdfzp})) and compare them to those obtained from a ``classical" and a PIMD simulation, both for an identical system. By ``classical'' we mean a standard $NVT$ MD simulation at $T$=300 $K$, with the use of a NH thermostat (without any GLE). 
 The PIMD simulation was  performed using the same method as described in Ref.~(\onlinecite{Ramirez2010a}), using a polymer ring of $P$=32 beads. Fig. (\ref{rdfzp}) clearly demonstrates that using NHGLE (with NHGLE-C parameters, see Table~(\ref{tab:pterms})) we can equilibrate modes to their average zero point temperature and the obtained 
structures are in agreement with PIMD results. This is the most important result of our work. We simultaneously reproduce  not only the O-O, but
more importantly the  O-H and H-H correlation functions. These results show that our hypothesis of setting each mode to their corresponding zero point
temperature is good enough to capture all the quantum effects on the structural properties of water. All the three pair correlation functions are reproduced with good accuracy within purely classical simulation.

\section{Analysis and Discussion}
In this section we present analysis of our results of selective mode thermostating of liquid water. Based on our results we are in a position to provide deeper insight into the role of individual modes contributing to the overall structure of liquid water.

\subsection{Vibrational spectrum}

We study the changes on the vibrational spectrum induced by the new distribution of normal mode temperatures. 
Doing this we can evaluate the intrinsic anharmonicities of liquid water (which strongly depend on the underlying model) at the correct zero-point temperature distribution.

\begin{center}
\begin{figure}[htb!]
\includegraphics[width=9.1cm,height=5.4cm]{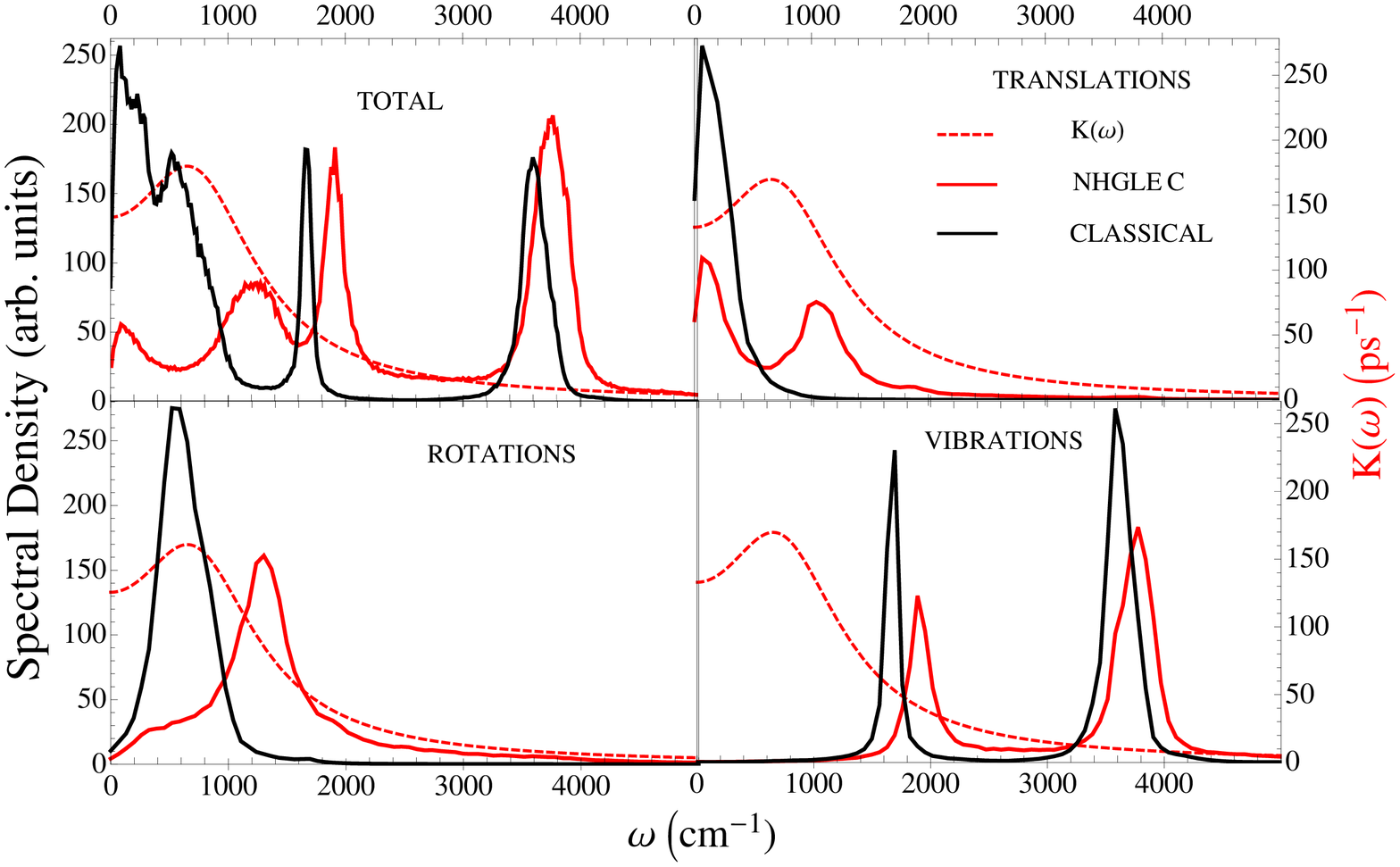}
\caption{(Color online) Spectral density plots decomposed into different vibrational contributions. Total spectrum (top left panel).
Projection onto translations (top right), rotations (bottom left) and vibrations (bottom right). Solid black line, NVT
 simulation at $T$=300 K. Red solid line, NHGLE C (see Table~(\ref{tab:pterms})) simulation.
The red dashed line shows the frequency dependent friction profile used in NHGLE C. The axes ticks on the left edge corresponds to the spectral density. The axes ticks on the right edge of the frame corresponds to $K(\omega)$ in units of $ps^{-1}$.}
\label{vvzp}
\end{figure}
\end{center}

In order to accurately evaluate how each region of the vibrational spectrum changes, we have obtained
the spectral density projected onto translational, rotational and vibrational modes separately. The procedure is straightforward once the cartesian coordinates
are projected onto the normal mode coordinates. The spectral density is obtained by computing the Fourier transform of the velocity-velocity
autocorrelation function within each independent normal mode subspace. Results are presented in Fig. (\ref{vvzp}). The figure shows both the changes
to the total spectrum (top left panel), and to the translations (top right), rotations (bottom left) and vibrations (bottom right).

The partition of the spectrum helps on identifying which are the modes that undergo major frequency shifts upon addition of quantum effects.
The largest, and more interesting changes occur in the translational and rotational regions of the spectrum. While the lowest frequency translational motions remain unaffected,
those modes with classical frequencies above $\approx$200 cm$^{-1}$ are strongly blue shifted. These are modes associated to the Hbond stretching.
The rotational peak also undergoes a large blue shift, although some of the modes also remain unaffected. The shifted rotational modes most likely correspond to
those associated to Hbond bending modes. In both cases, the shift is a consequence of the large temperature NHGLE imposes on the intramolecular vibrational modes.
Indeed, as seen in Table~(\ref{tab:pterms}), label ``NHGLE C" and in Fig.~\ref{tempall} (d), the net $T$ of the vibrational modes is ~2200 K.
This large blue shift confirms the strong coupling between inter and intra molecular modes, or a strong intrinsic anharmonicity.
This shift also modifies their corresponding zero point temperature. To account for this zero point shift, our NHGLE parameters are tuned
to equilibrate intermolecular modes to 600 K (trans) 800 K (rotations) whereas intramolecular modes are equilibrated at 2200 K.
Interestingly, this method, with a very different methodology and mathematical framework, is an alternative
to include the correct level of anharmonicities to the method proposed by Hardy {\it et al.}\cite{Hardy98} using
a combination of classical molecular dynamics and the quasiharmonic approximation.

Our scheme is particularly suited to extract dynamical information. The method ensure the absence of zero-point leakage\cite{pimdspectrum,zpleakage}, as seen in the conservation of mode-projected temperatures through the simulation length. 

\subsection{Emergence of water structure from stretching modes}\label{sec:540k}
The results for zero point simulation using NHGLE can also be interpreted from a different view point. In this section we compare classical simulation done at 600 K to our effective zero point simulation done at  600 K (translations), 800 K (rotations) and 2200 K (vibrations) (``NHGLE C" in Table~(\ref{tab:pterms})).

\begin{center}
\begin{figure}[htb!]
\includegraphics[width=8.2cm,height=5cm]{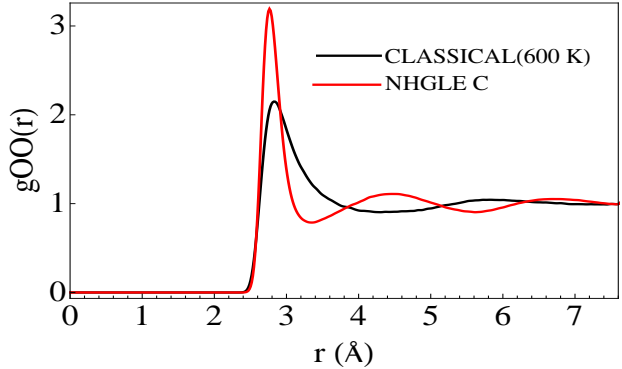}\\
\caption{O-O Radial distribution function. Red line: NHGLE C (see  Table~(\ref{tab:pterms})) with
a temperature distribution of: 600 K translations, 800 K rotations and 2200 K vibrations. Black line: NVT simulation
 equilibrated at 600 K.}
\label{540}
\end{figure}
\end{center}
Fig. (\ref{540}) shows that for a classical simulation equilibrated at 600 K the liquid water structure is completely lost, as expected for such a high temperature. Indeed
at this $T$ the liquid is at a supercritical state, because we do not allow the volume of our simulation cell to change. The O-O rdf in red is a liquid water structure that emerges completely due to the higher temperature imposed on the stretching modes. This comparison clearly points out that zero point temperature of O-H stretching modes plays a dominant role in the structuring of liquid water.

\subsection{Role of individual modes from PIMD simulations.}\label{sec:pimd}

So far in this work we have demonstrated how quantum effects corresponding to individual modes effect the overall structure of liquid water. In this section we try to extract this information from PIMD simulations. Recent work of Habershon {\it et al.}\cite{Manolopoulos09} established the idea of competing quantum effects by understanding the role of intramolecular stretching modes. They observe reduced quantum to classical ratio of diffusion coefficient when the O-H stretching is allowed. This is due
to the anharmonicity of the stretching mode, which couples it to the rotational and translational modes. This is in agreement with our results. However in this work we aim to understand this effect in terms of the zero point temperature of competing modes. In PIMD simulations, we map the zero point temperature on individual modes to the number of beads.  The quantum limit is achieved with $P\rightarrow \infty$, $P$ being the number of beads employed for the discretization of the path integral.
 A PIMD simulation with finite P at temperature T implies a high-temperature approximation, i.e. the partition function of the system is considered to be classical at a higher temperature given by the product $PT$. Usually the value of $P$ is chosen so that the product $PT$ is several times larger than the zero point temperature of the highest frequency.

\begin{center}
\begin{figure}[htb!]
\includegraphics[width=8.2cm,height=5cm]{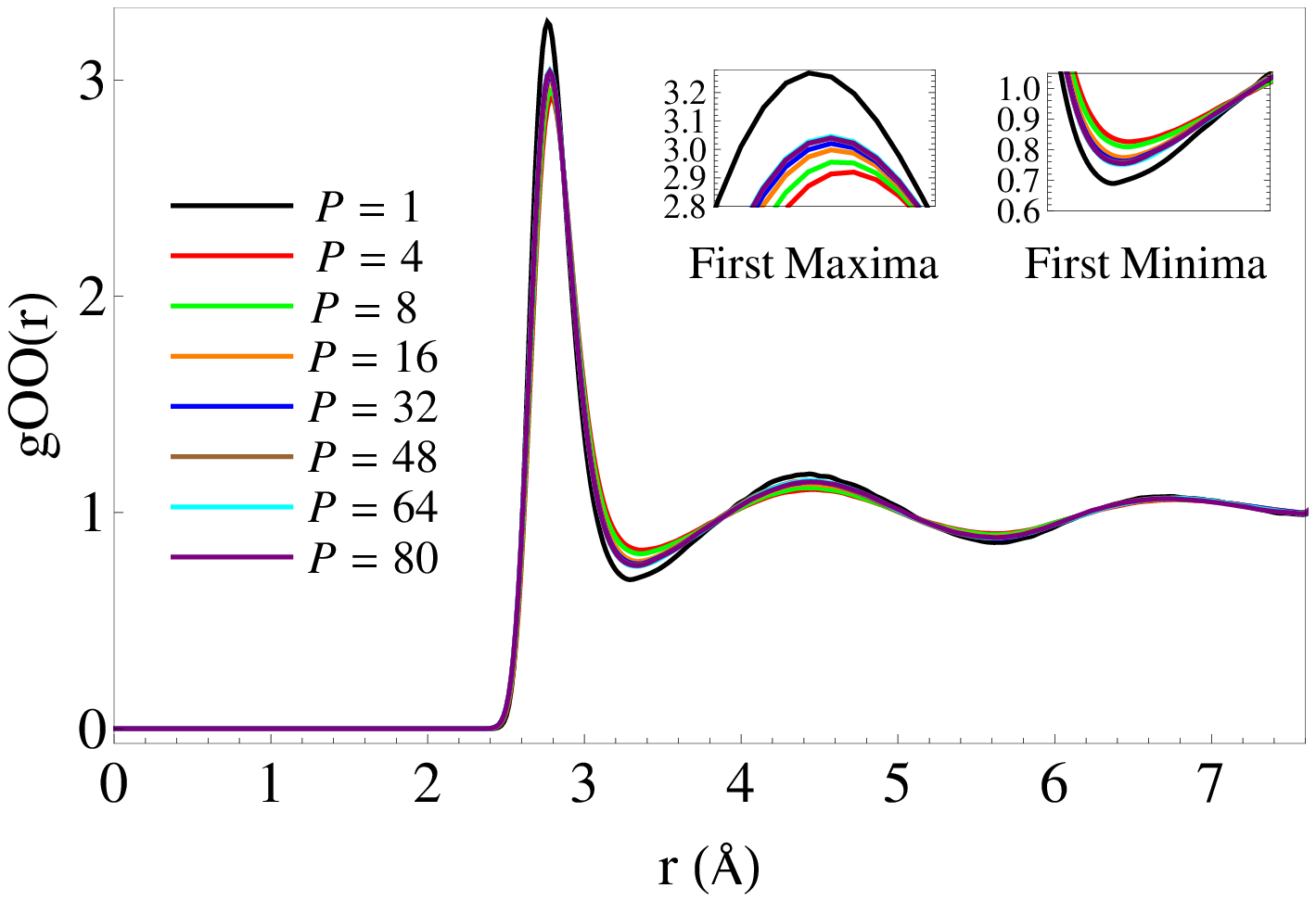}\\
\vspace{0.2cm}
\includegraphics[width=8.2cm,height=5cm]{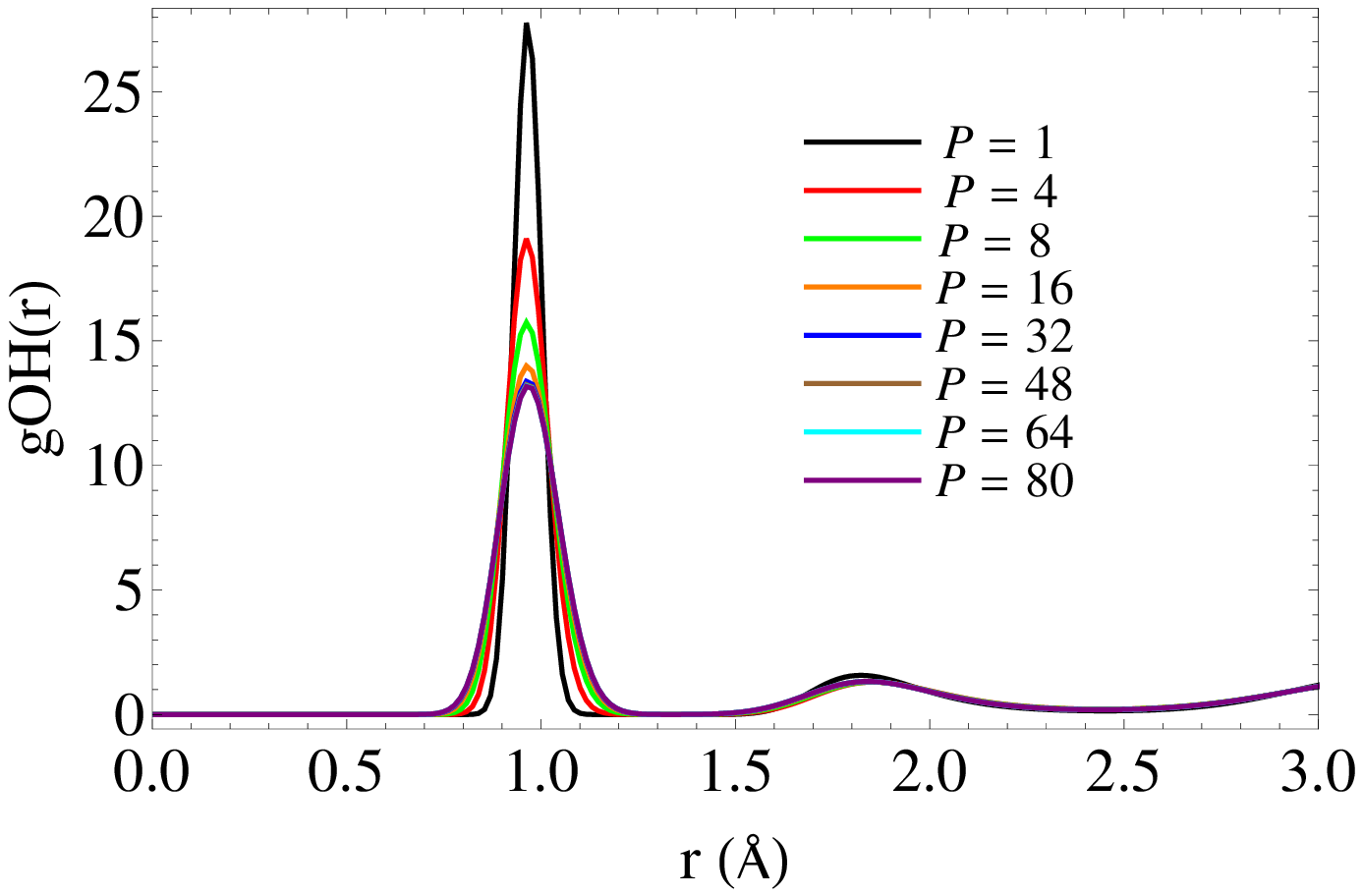}
\caption{(Color online)Radial distribution function plots for PIMD simulations as a function of the number of beads. Top: O-O rdf. The two insets
are zoomed into the first peak and the first minima. Bottom: O-H rdf}
\label{rdfpimd}
\end{figure}
\end{center}

We study the structure of liquid water as a function of $P$. We plot the rdf obtained from PIMD for O-O  and O-H pairs, for a 256 molecules water simulation of 250 ps length and density 0.997 $g/cm^{3}$   (see Fig. \ref{rdfpimd}).

\begin{center}
\begin{figure}[htb!]
\includegraphics[width=8.2cm,height=5cm]{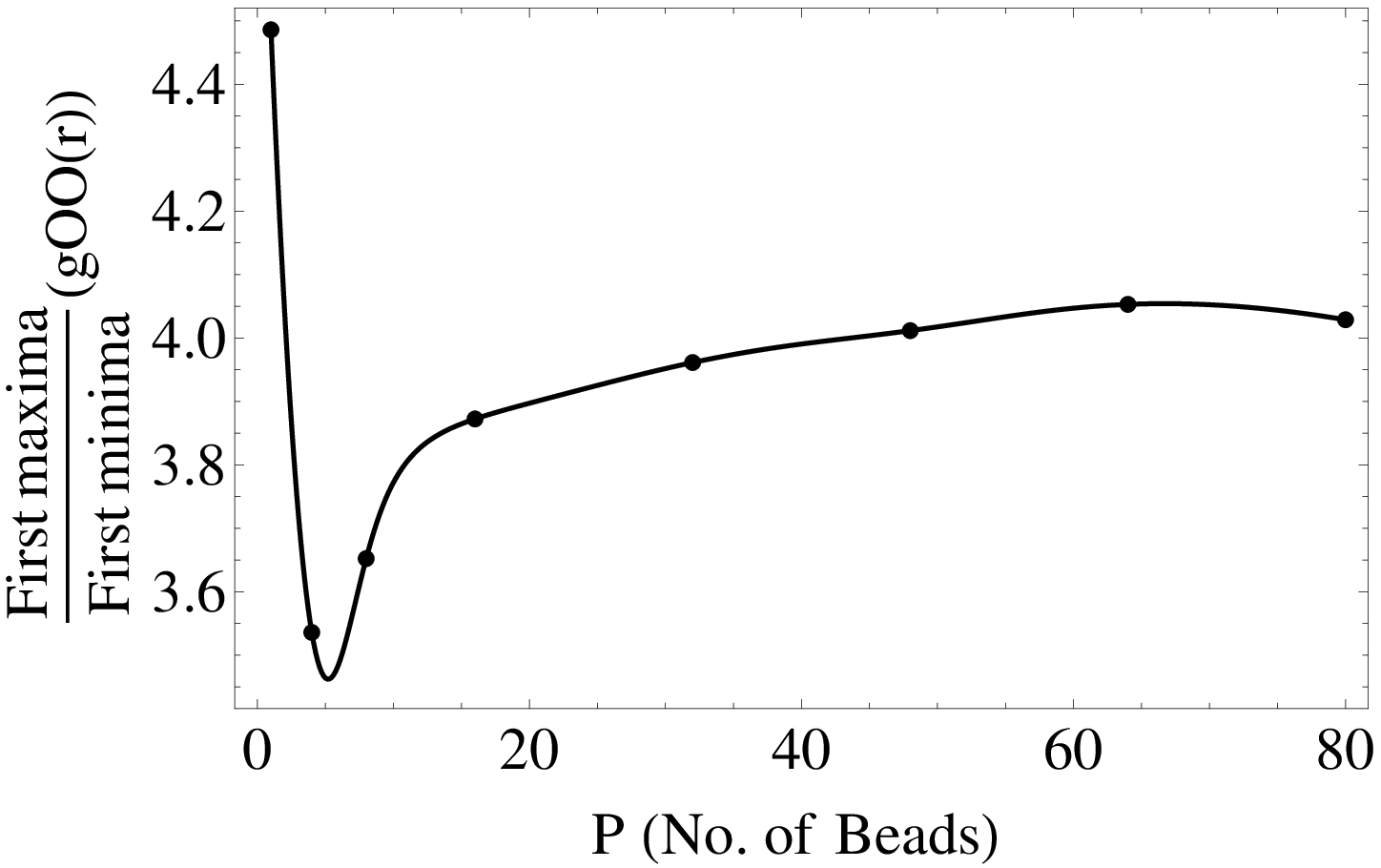}\\
\vspace{1cm}
\includegraphics[width=8.2cm,height=5cm]{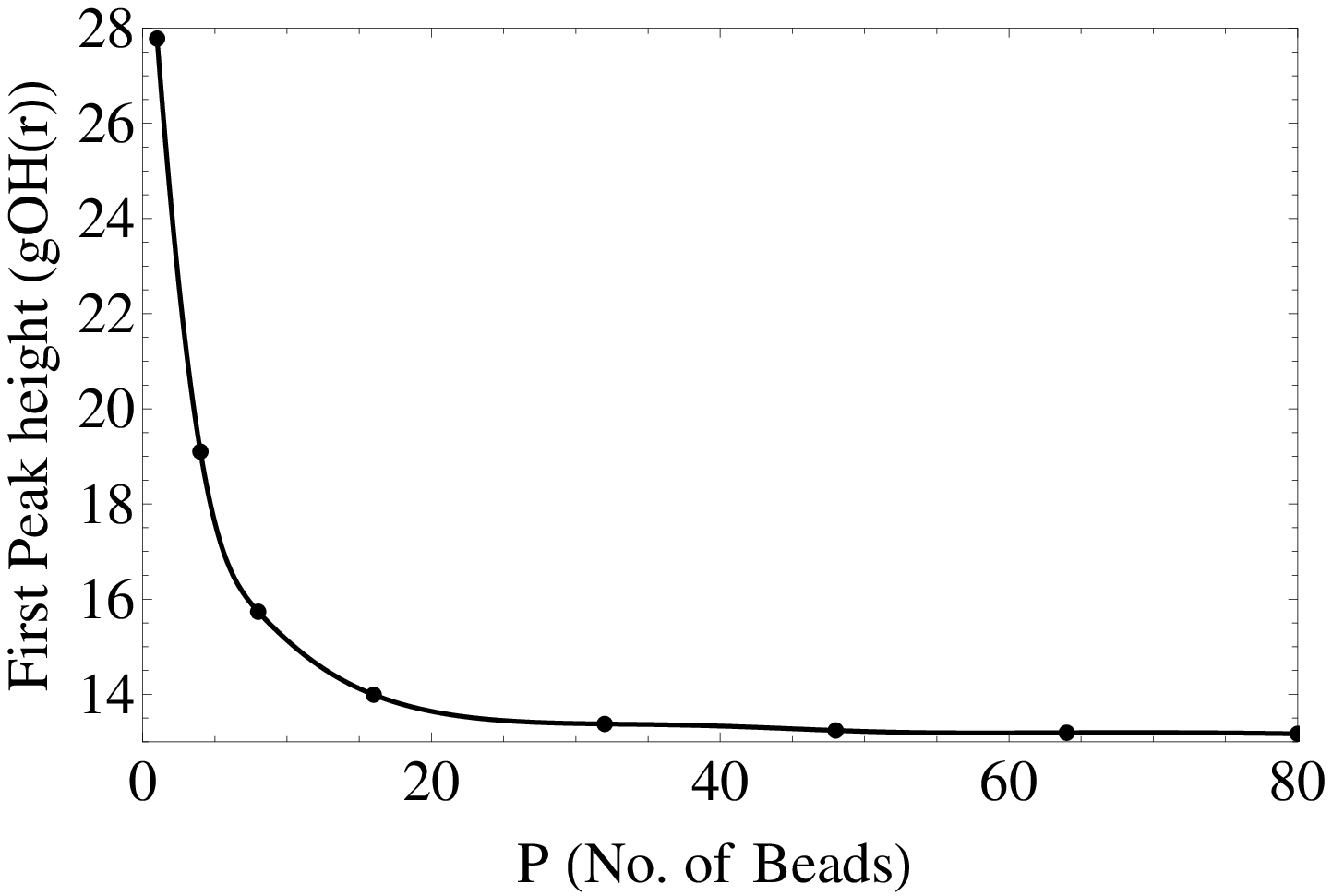}
\caption{Order parameter as a function of the number of beads in PIMD simulations. Top: ratio of the height
 of first maxima to first minima. Bottom: height of first peak of the O-H rdf.
The points are the actual data and the solid line is a fourth order interpolation. }
\label{pimd_correl}
\end{figure}
\end{center}

Based on the plots in Fig. (\ref{rdfpimd}) we can analyze how the zero point effects of each mode influences the structure of liquid water. To understand the structure we introduce
 an order parameter which is the ratio of the first maxima to the first minima of the O-O rdf. The increasing value of this ratio is related to the increasing structure of liquid water. We plot this ratio as a function of $P$. For the O-H rdf we plot the height of the first peak as a function of $P$ (see Fig.~(\ref{pimd_correl})).

In Fig. (\ref{pimd_correl}) we observe that for $P\le10$, i.e. for $PT\le 3000 $K  goo(r) looses structure compared to the classical (P=1 bead) case. This can be attributed to the fact that using only a few beads, the zero point temperature of intermolecular modes is well captured and they tend to soften the liquid structure. However for small $P$
 the PIMD simulation is not able to describe the zero point vibration of the O-H stretching modes. We have already mentioned that the  temperature PT must be several times larger that the zero point temperature for a reasonable approximation of the quantum limit in a PIMD simulation.. As we further increase no. of beads we see that goo(r) starts gaining structure. This gain in structure is due to the better and better description of zero point effects of intramolecular modes. This gain in structure saturates when the intramolecular modes are kept close to their zero point temperature. Hence using $P$ as a control parameter for quantum effects, we can understand the competition between intra and intermolecular modes. We also see a crossover from unstructuring to structuring in goo(r) (Fig. (\ref{rdfpimd})) as we increase the no. of beads. A model with too little anticorrelation as the one used here\cite{Pamuk2012}, will not
structure above the classical level, independently of the number of beads used.

\section{Conclusions}
In this work, we have combined NH and GLE thermostats to create a powerful selective mode thermostating scheme, that we coin NHGLE.
The fundamental idea behind this thermostat is to be able to simulate the quantum mechanical nature of the nuclei by
setting the temperature of the normal modes they participate in to their corresponding zero-point temperature.
 Using NHGLE we were able to achieve this goal, ensuring the maintenance of the NHGLE-imposed non-equilibrium temperature distribution
for any simulation time length. The GLE noise is suppressed  by setting the GLE temperature close to zero, making
the thermostat dynamics deterministic.
We have applied NHGLE to a flexible force field model of water (q-TIP4P/F).
Our results show that the structure of liquid water changes when intramolecular modes are equilibrated to a different temperature than that of intermolecular modes.
We showed that equilibrating vibrations at slightly lower temperature than other modes results in the loss of structure in liquid water. Conversely water is more structured if we equilibrate vibrations to slightly higher temperature. This simple exercise verifies the well known anticorrelation effect \cite{Pamuk2012} in terms of the temperature of individual modes.
Simply changing NHGLE parameters we can set all modes to their corresponding zero point temperature. 
Our zero point estimates of translations and rotations prove to be correct as NHGLE simulations reproduce O-O, O-H and H-H rdf obtained from PIMD simulation with 32  beads.
Finally, we have analyzed selective mode behavior as a function of their zero point temperature in PIMD.
To this aim we showed that by a systematic increase of the number of beads  $P$ in  PIMD simulations, the different vibrational modes of water can be successively  tuned from a classical to a quantum limit.
The height of first peak of the O-H rdf continuously decreased as a function of $P$.
This effect is similar to adding more zero point temperature on vibrations.
For $P\le10$, there is drastic softening of long range structure as seen in the O-O rdf.
 This is the case when the temperature of low energy modes are close to their zero point temperature but the vibration temperature is still far away from its zero point.
For $P>10$ we observe an increase of structure in the O-O rdf that asymptotically saturates with $P$.
This is a consequence of intramolecular modes equilibrated to their zero point temperature.
 In summary, we successfully introduced a selective mode thermostating scheme (NHGLE), which can be easily tuned to include quantum zero point effects, producing results in good agreement with PIMD.
We used it to study how the structure of liquid water responds to different temperatures, unveiling the existence of large intrinsic anharmonicities in the vibrational modes
of liquid water.
In conclusion, this method represents an efficient and simple alternative to other proposed solutions of how to perform classical molecular dynamics simulations with quantum accuracy\cite{Poma10}. Its performance in combination with ab-initio molecular dynamics simulations, where the force field dependence aspects can be filtered out,  needs to be evaluated. This will be the subject of a future publication.
\acknowledgments

We thanks Jose M. Soler for providing us with an early version of the code to perform the
projections onto normal modes and for very useful discussions and comments.
This work was partially supported by DOE
award number DE-FG02-09ER16052 (SG)  and by
DOE Early Career award DE-SC0003871
(MVFS).
The work from RR in Madrid is supported by
Ministerio de Ciencia e Innovaci\'on (Spain)
through Grant No. FIS2012-31713, and by Comunidad Aut\'onoma de Madrid
through project MODELICO-CM/S2009ESP-1691.

\appendix
 \section{Microscopic interpretation of NHGLE dynamics}\label{sec:glemodel}
In this section we start with the microscopic model for NHGLE thermostat starting from a system-bath coupling model. The full extended Hamiltonian of the system can be written as,
\be
H_{total}=H_{sys}(\frac{p_i}{s},q_i)+H_{NH}(p_{s},s)+H_{GLE}(p_{i,x_k},x_{i,k})
\ee
Where,
\bea
H_{sys}&=&\sum^{3N}_{i=1}\frac{p^{2}_{i}}{2M_is^2}+V(q_1....q_{3N})\\
H_{NH}&=&\frac{p^{2}_{s}}{2Q}+(3N+1)k_BT_{NH}\log s\nonumber\\
H_{GLE}&=&\sum^{3N}_{i=1}H^i_{GLE}\\
H^i_{GLE}&=&\sum^{g}_{k=1}\left(\frac{p^{2}_{i,x_k}}{2m}+\frac{1}{2}m\omega_k^2\Big(x_{i,k}+\frac{\beta\ q_i}{m\omega_k^2}\Big)^2\right)\nonumber\\
\eea
$s$ is the parameter that modifies the effective dynamics and enforces constant temperature ensemble on the rescaled momenta ($\frac{p}{s}$). $Q$ is the NH mass. This rescaled dynamics is also coupled to a harmonic bath that enforces generalized Langevin dynamics. This is achieved by coupling each system degree of freedom to $g$ harmonic oscillators of mass $m$ and frequency $\omega_k$. $\beta$ is the coupling strength of the oscillator to the system degree of freedom. $p_{i,x_k}$ is the momentum conjugate to the kth oscillator position $x_{i,k}$ (index i corresponds to the system degree of freedom). Hamilton Jacobi equations for the total dynamics for the system degrees of freedom can be written as,
 \bea
 \dot{q}_i&=&\frac{p_i}{M_is^2}\\
 \dot{p}_i&=&-\frac{\partial V(\bf{q})}{\partial q_i}-\sum_k \frac{\beta^2}{m\omega_k^2} q_i-\sum_k\beta x_{i,k}
  \label{eq:ngle}
\eea
The dynamics of NH degrees of freedom is given as,
\bea
 \dot{s}&=& \frac{p_{s}}{Q}\nonumber\\
  \dot{p}_{s}&=&\sum^{3N}_i\frac{p^{2}_i}{M_is^3}-\frac{(3N+1) k_B T_{NH}}{s}
  \label{eq:nose}
\eea
We can further simplify NH dynamics in Eq.~(\ref{eq:nose}) by rescaling time and momenta. We perform $dt\rightarrow\frac{dt}{s}$ in Eq.~(\ref{eq:ngle}). The resulting equations can be written in terms of the rescaled momenta $p_i\rightarrow\frac{p_i}{s}$
\bea
 \dot{\eta}&=& \frac{p_{\eta}}{Q}\nonumber\\
 \dot{p}_{\eta}&=&\Big[\sum^{N}_i\frac{p^{2}_i}{M_i}-(3N+1) k_B T_{NH}\Big]\nonumber\\
\eea
Where $\eta=\log s$.

The dynamics of GLE bath degrees of freedom is given by.
  \bea
 \dot{x}_{i,k}&=&\frac{p_{i,x_k}}{m_k}\\
 \dot{p}_{i,x_k}&=&-\beta q_i-m \omega_k^2x_{i,k}
  \label{eq:glebath}
\eea
The above equations for the GLE bath can be solved exactly and can be substituted in Eq.~(\ref{eq:ngle}). The resulting system's dynamics in presence of NH chains and the  GLE bath\cite{fordmazur,zwanzig1973} can be written in the following form,
\begin{eqnarray}
\dot{q}_i & = & \frac{p_i}{M_i}\label{eq:1dgle-1}\\
\dot{p}_i & = & -\frac{\partial V}{\partial q_i}-\int_{-\infty}^{t}K(t-t')p_i(t')dt'+\zeta(t)-p_i\frac{p_{\eta}}{Q}\nonumber\\\label{eq:1dgle-2}
\end{eqnarray}
where the last term is the NH term that is coupled to the system. The memory kernel $K(t)$ has an exact expression in terms of the bath parameters, \be K(t)=\sum_k\left(\frac{\beta}{\sqrt{m}\omega_k}\right)^2\cos(\omega_kt)\label{bathkernel}\ee
The `random' force $\zeta(t)$ can also be completely determined in terms of the bath degrees of freedom. $\zeta(t)$ is connected to the memory kernel through the FD theorem as $\langle\zeta(t)\zeta(t')\rangle=k_BT_{GLE}\ K(t-t')$ with $\langle\zeta(t) \rangle=0$ (The existence of the FD relation with respect to the GLE bath is an approximation under the assumption that we can define a local equilibrium condition. $T_{GLE}$ is defined as the temperature that would be enforced on the system by the GLE thermostat in the absence of the NH thermostat.).  

Note that the temperature enforced by the FD condition is different from the NH temperature.
We can rewrite Eq.~(\ref{eq:1dgle-2}) by absorbing the NH term into the time integral.
\bea
\dot{p}_i & = & -\frac{\partial V}{\partial q_i}-\int_{-\infty}^{t}\tilde{K}(t,t')p_i(t')dt'+\sqrt{2M_ik_BT_{GLE}}\tilde{\zeta}(t)\nonumber\\\label{eq:nosekernel}
\eea
Where we have defined,
\bea
\tilde{K}(t,t')&=& K(t-t')+\delta(t-t')\frac{p_{\eta}(t')}{Q}\ ,\ \langle\tilde{\zeta}(t)\tilde{\zeta}(0)\rangle=K(t)
\eea
Now consider Eq.~(\ref{eq:nosekernel}) in the case when $T_{GLE}\rightarrow 0$,
\be
\dot{p}_i=-\frac{\partial V}{\partial q_i}-\int_{-\infty}^{t}\tilde{K}(t,t')p_i(t')dt'
\label{eq:nonoise}
\ee
In this case the noise term is completely suppressed and we have a GLE dynamics that involves only
a nonlocal friction profile and the deterministic NH term that provides constant temperature $T_{NH}$ to the system. Schematically, a more precise statement would be that we subject our system to two thermostats. The GLE thermostat on its own would have enforced $T_{system}=T_{GLE}=0K$. The NH thermostat on its own would have enforced $T_{system}=T_{NH}$. Interesting non-equilibrium (frequency dependent) temperature distribution is a result of interplay between these thermostats. The validity of our approach is independent  of the microscopic Hamiltonian model. We can always treat the non-local kernel $K(t)$ defined in Eq.~(\ref{eq:1dgle-2}) as a modification to the NH dynamics while  coupling to the system degrees of freedom. This modification enables frequency dependent coupling of the NH to the system degrees of freedom which is entirely captured in Eq.~(\ref{eq:nonoise}).

The conserved energy for this modified dynamics can be written as,
\bea
H'&=&H_{sys}(p,q)+\frac{p^2_{\eta}}{2Q}+(3N+1)k_B T_{NH}\eta+\sum_i\Delta KE_i\nonumber\\
\label{eq:consv}
\eea
The last term in the above equation is the change in kinetic energy for each GLE action summed over all past trajectories\cite{bussi,ceriotti3}. In this work we achieve frequency dependent equilibration as a result of the competition between the NH dynamics and the damping action of nonlocal friction profile with suppressed noise. %
\subsection{Delta-like memory kernels}
The friction term in Eq.~(\ref{eq:1dgle-2}) is linear in the system momentum, and the friction coefficient $K(t)$ is a simple function of the frequencies $\omega_k$ and the coupling constants $\frac{\beta}{\sqrt{m}}$. This generalized Langevin equation is exact and its validity is not restricted to small departures from thermal equilibrium. Now we consider the case of infinite number of oscillators with continuous distribution of frequencies $\omega_k$. In this limit the summation in Eq.~(\ref{bathkernel}) can be replaced by an integral with some distribution function ($\sum \rightarrow N\int d\omega g(\omega)$).
\be K(t)=N\int d\omega\ g(\omega)\left(\frac{\beta}{\sqrt{m}\omega}\right)^2\cos(\omega\ t)\label{contbathkernel}\ee
Since we have control over the bath degrees of freedom we choose the coupling constant to be, $\frac{\beta}{\sqrt{m}}=\sqrt{\frac{\gamma(\Delta\omega^{2}+\omega_{0}^{2})}{N}}$. We are free to choose the distribution of frequencies of the oscillators to enforce desired memory kernel on the system. To design delta  like memory kernels $K(t)$ introduced in Ref.~(\onlinecite{ceriotti4}), we choose the frequency distribution function of the oscillators to be,
\be
g(\omega)=\frac{\omega^2}{\Delta\omega^{2}+(\omega-\omega_{0})^{2}}
\ee
The above distribution of the continuum oscillator frequencies results in the effective delta like friction profile of Ref.~(\onlinecite{ceriotti4}) which is the essential component of frequency dependent thermostating scheme. The memory kernel obtained is given by,
\begin{eqnarray}
K(\omega) & = & \gamma\Bigg(\frac{\Delta\omega^{2}+\omega_{0}^{2}}{\Delta\omega}\Bigg)\Bigg(\frac{\Delta\omega}{\Delta\omega^{2}+(\omega-\omega_{0})^{2}}\nonumber\\
 & + & \frac{\Delta\omega}{\Delta\omega^{2}+(\omega+\omega_{0})^{2}}\Bigg)\label{kernel}\\
K(t) & = & \gamma\Bigg(\frac{\Delta\omega^{2}+\omega_{0}^{2}}{\Delta\omega}\Bigg)e^{-\mid t\mid\Delta\omega}\cos\omega_{0}t
\end{eqnarray}
In the memory kernel defined above, $\gamma$ is the friction coefficient (or coupling strength to the harmonic bath). $\omega_{0}$ is the frequency at which the delta shaped memory kernel has
maximum friction value or maximum strength of coupling. $\Delta\omega$ is the width of the friction profile. Notice that all the parameters related
to the GLE thermostat are completely independent of the force field parameters. All we need to know is the position of the peaks of modes that we can
obtain from the spectral density of the system in consideration.
Non-Markovian  dynamics can be mapped to Markovian dynamics in higher dimensional space\cite{ceriotti3} by adding auxiliary momentum degrees of freedom\cite{note1}. The modified higher dimensional dynamics can be implemented in the form of following dynamical equations,
\vspace{-0.1cm}
\begin{eqnarray}
\dot{q} & = & \frac{p}{M}\\
\left(\begin{array}{c}
\dot{p}\\
{\bf \dot{s}}\end{array}\right) & = & \left(\begin{array}{c}
-\frac{\partial V}{\partial x}\\
0\end{array}\right)-{\bf A}\left(\begin{array}{c}
p\\
{\bf s}\end{array}\right)+{\bf B}\xi(t)\end{eqnarray}

Matrices ${\bf A}$ and ${\bf B}$ are the drift and diffusion matrices respectively. $\xi$ is the uncorrelated Markovian noise, and ${\bf s}$ is the vector of additional momentum degrees of freedom. The drift and diffusion matrices may be constrained by the generalized FD
theorem
\begin{eqnarray}
{\bf BB^{T}}&=&Mk_{B}T_{GLE}({\bf A+A^{T}})\\
{\bf BB^{T}}&=&0\ \ \ (T_{GLE}=0)
\label{eq:FD}
\end{eqnarray}

The matrix ${\bf A}$ has the form $\left(\begin{array}{cc}
a_{pp} & {\bf a_{ps}}\\
{\bf a_{ps}^{T}} & {\bf a}\end{array}\right)$. One can obtain functional form of the memory kernel from the matrix
${\bf A}$.
\vspace{-0.1cm}
\begin{equation}
K(t)=2a_{pp}\delta(t)-{\bf a_{ps}}e^{-\mid t\mid{\bf a}}{\bf a_{ps}^{T}}\label{eq:memorykernel}
\end{equation}
The drift matrix ${\bf {\bf A}}$ for that produces delta like friction profile in Eq.~(\ref{kernel}) is given by,
\be
{\bf A}=\left(\begin{array}{ccc}
0 & \sqrt{\gamma(\frac{\Delta\omega^{2}+\omega_{0}^{2}}{2\Delta\omega})} & \sqrt{\gamma(\frac{\Delta\omega^{2}+\omega_{0}^{2}}{2\Delta\omega})}\\
-\sqrt{\gamma(\frac{\Delta\omega^{2}+\omega_{0}^{2}}{2\Delta\omega})} & \Delta\omega & \omega_{0}\\
-\sqrt{\gamma(\frac{\Delta\omega^{2}+\omega_{0}^{2}}{2\Delta\omega})} & -\omega_{0} & \Delta\omega\end{array}\right)
\label{delta-like}
\ee


\begin{thebibliography}{0}%
\makeatletter
\providecommand \@ifxundefined [1]{%
 \@ifx{#1\undefined}
}%
\providecommand \@ifnum [1]{%
 \ifnum #1\expandafter \@firstoftwo
 \else \expandafter \@secondoftwo
 \fi
}%
\providecommand \@ifx [1]{%
 \ifx #1\expandafter \@firstoftwo
 \else \expandafter \@secondoftwo
 \fi
}%
\providecommand \natexlab [1]{#1}%
\providecommand \enquote  [1]{``#1''}%
\providecommand \bibnamefont  [1]{#1}%
\providecommand \bibfnamefont [1]{#1}%
\providecommand \citenamefont [1]{#1}%
\providecommand \href@noop [0]{\@secondoftwo}%
\providecommand \href [0]{\begingroup \@sanitize@url \@href}%
\providecommand \@href[1]{\@@startlink{#1}\@@href}%
\providecommand \@@href[1]{\endgroup#1\@@endlink}%
\providecommand \@sanitize@url [0]{\catcode `\\12\catcode `\$12\catcode
  `\&12\catcode `\#12\catcode `\^12\catcode `\_12\catcode `\%12\relax}%
\providecommand \@@startlink[1]{}%
\providecommand \@@endlink[0]{}%
\providecommand \url  [0]{\begingroup\@sanitize@url \@url }%
\providecommand \@url [1]{\endgroup\@href {#1}{\urlprefix }}%
\providecommand \urlprefix  [0]{URL }%
\providecommand \Eprint [0]{\href }%
\providecommand \doibase [0]{http://dx.doi.org/}%
\providecommand \selectlanguage [0]{\@gobble}%
\providecommand \bibinfo  [0]{\@secondoftwo}%
\providecommand \bibfield  [0]{\@secondoftwo}%
\providecommand \translation [1]{[#1]}%
\providecommand \BibitemOpen [0]{}%
\providecommand \bibitemStop [0]{}%
\providecommand \bibitemNoStop [0]{.\EOS\space}%
\providecommand \EOS [0]{\spacefactor3000\relax}%
\providecommand \BibitemShut  [1]{\csname bibitem#1\endcsname}%
\let\auto@bib@innerbib\@empty
\end{thebibliography}%


\begin{thebibliography}{22}
\expandafter\ifx\csname natexlab\endcsname\relax\def\natexlab#1{#1}\fi
\expandafter\ifx\csname bibnamefont\endcsname\relax
  \def\bibnamefont#1{#1}\fi
\expandafter\ifx\csname bibfnamefont\endcsname\relax
  \def\bibfnamefont#1{#1}\fi
\expandafter\ifx\csname citenamefont\endcsname\relax
  \def\citenamefont#1{#1}\fi
\expandafter\ifx\csname url\endcsname\relax
  \def\url#1{\texttt{#1}}\fi
\expandafter\ifx\csname urlprefix\endcsname\relax\def\urlprefix{URL }\fi
\providecommand{\bibinfo}[2]{#2}
\providecommand{\eprint}[2][]{\url{#2}}

\bibitem[{\citenamefont{Habershon et~al.}(2009)\citenamefont{Habershon,
  Markland, and Manolopoulos}}]{Manolopoulos09}
\bibinfo{author}{\bibfnamefont{S.}~\bibnamefont{Habershon}},
  \bibinfo{author}{\bibfnamefont{T.~E.} \bibnamefont{Markland}},
  \bibnamefont{and} \bibinfo{author}{\bibfnamefont{D.~E.}
  \bibnamefont{Manolopoulos}}, \bibinfo{journal}{J. Chem. Phys.} \textbf{\bibinfo{volume}{131}}, \bibinfo{eid}{024501}
  (pages~\bibinfo{numpages}{11}) (\bibinfo{year}{2009}).
  
  \bibitem [{\citenamefont {Vega}\ \emph {et~al.}(2010)\citenamefont {Vega},
  \citenamefont {Conde}, \citenamefont {McBride}, \citenamefont {Abascal},
  \citenamefont {Noya}, \citenamefont {Ramirez},\ and\ \citenamefont
  {Ses\'{e}}}]{vega10}%
  \BibitemOpen
  \bibfield  {author} {\bibinfo {author} {\bibfnamefont {C.}~\bibnamefont
  {Vega}}, \bibinfo {author} {\bibfnamefont {M.~M.}\ \bibnamefont {Conde}},
  \bibinfo {author} {\bibfnamefont {C.}~\bibnamefont {McBride}}, \bibinfo
  {author} {\bibfnamefont {J.~L.~F.}\ \bibnamefont {Abascal}}, \bibinfo
  {author} {\bibfnamefont {E.~G.}\ \bibnamefont {Noya}}, \bibinfo {author}
  {\bibfnamefont {R.}~\bibnamefont {Ramirez}}, \ and\ \bibinfo {author}
  {\bibfnamefont {L.~M.}\ \bibnamefont {Ses\'{e}}},\ }\href {\doibase
  10.1063/1.3298879} {\bibfield  {journal} {\bibinfo  {journal} {The Journal of
  Chemical Physics}\ }\textbf {\bibinfo {volume} {132}},\ \bibinfo {eid}
  {046101} (\bibinfo {year} {2010})}\BibitemShut {NoStop}%

\bibitem [{\citenamefont {Paesani}\ \emph {et~al.}(2010)\citenamefont
  {Paesani}, \citenamefont {Yoo}, \citenamefont {Bakker},\ and\ \citenamefont
  {Xantheas}}]{paesani10}%
  \BibitemOpen
  \bibfield  {author} {\bibinfo {author} {\bibfnamefont {F.}~\bibnamefont
  {Paesani}}, \bibinfo {author} {\bibfnamefont {S.}~\bibnamefont {Yoo}},
  \bibinfo {author} {\bibfnamefont {H.~J.}\ \bibnamefont {Bakker}}, \ and\
  \bibinfo {author} {\bibfnamefont {S.~S.}\ \bibnamefont {Xantheas}},\ }
  {\bibfield  {journal} {\bibinfo  {journal} {The
  Journal of Physical Chemistry Letters}\ }\textbf {\bibinfo {volume} {1}},\
  \bibinfo {pages} {2316} (\bibinfo {year} {2010})}.
  
 \bibitem [{\citenamefont {Gonz\'alez}\ \emph {et~al.}(2010)\citenamefont
  {Gonz\'alez}, \citenamefont {Noya}, \citenamefont {Vega},\ and\ \citenamefont
  {Ses\'e}}]{briesta10}%
  \BibitemOpen
  \bibfield  {author} {\bibinfo {author} {\bibfnamefont {B.~S.}\ \bibnamefont
  {Gonz\'alez}}, \bibinfo {author} {\bibfnamefont {E.~G.}\ \bibnamefont
  {Noya}}, \bibinfo {author} {\bibfnamefont {C.}~\bibnamefont {Vega}}, \ and\
  \bibinfo {author} {\bibfnamefont {L.~M.}\ \bibnamefont {Ses\'e}},\ }
  {\bibfield  {journal} {\bibinfo  {journal} {The
  Journal of Physical Chemistry B}\ }\textbf {\bibinfo {volume} {114}},\
  \bibinfo {pages} {2484} (\bibinfo {year} {2010})},\ \bibinfo {note} {pMID:
  20121175}.
  
\bibitem [{\citenamefont {Li}\ \emph {et~al.}(2011)\citenamefont {Li},
  \citenamefont {Walker},\ and\ \citenamefont {Michaelides}}]{zheng11}%
  \BibitemOpen
  \bibfield  {author} {\bibinfo {author} {\bibfnamefont {X.-Z.}\ \bibnamefont
  {Li}}, \bibinfo {author} {\bibfnamefont {B.}~\bibnamefont {Walker}}, \ and\
  \bibinfo {author} {\bibfnamefont {A.}~\bibnamefont {Michaelides}},\ }
   {\bibfield  {journal} {\bibinfo  {journal}
  {Proceedings of the National Academy of Sciences}\ }\textbf {\bibinfo
  {volume} {108}},\ \bibinfo {pages} {6369} (\bibinfo {year} {2011})}.


\bibitem[{\citenamefont{Ram\'{\i}rez and Herrero}(2010)}]{Ramirez2010a}
\bibinfo{author}{\bibfnamefont{R.}~\bibnamefont{Ram\'{\i}rez}}
  \bibnamefont{and} \bibinfo{author}{\bibfnamefont{C.~P.}
  \bibnamefont{Herrero}}, \bibinfo{journal}{J. Chem. Phys.}
  \textbf{\bibinfo{volume}{133}}, \bibinfo{eid}{144511}
  (pages~\bibinfo{numpages}{12}) (\bibinfo{year}{2010}).

\bibitem[{\citenamefont{Morrone and Car}(2008)}]{Car08}
\bibinfo{author}{\bibfnamefont{J.~A.} \bibnamefont{Morrone}} \bibnamefont{and}
  \bibinfo{author}{\bibfnamefont{R.}~\bibnamefont{Car}},
  \bibinfo{journal}{Phys. Rev. Lett.} \textbf{\bibinfo{volume}{101}},
  \bibinfo{pages}{017801} (\bibinfo{year}{2008}).

\bibitem[{\citenamefont{Herrero and Ram\'{\i}rez}(2011)}]{Ramirez11}
\bibinfo{author}{\bibfnamefont{C.~P.} \bibnamefont{Herrero}} \bibnamefont{and}
  \bibinfo{author}{\bibfnamefont{R.}~\bibnamefont{Ram\'{\i}rez}},
  \bibinfo{journal}{J. Chem. Phys.}
  \textbf{\bibinfo{volume}{134}}, \bibinfo{eid}{094510}
  (pages~\bibinfo{numpages}{10}) (\bibinfo{year}{2011}).

\bibitem[{\citenamefont{Pamuk et~al.}(2012)\citenamefont{Pamuk, Soler,
  Ram\'irez, Herrero, Stephens, Allen, and Fern\'andez-Serra}}]{Pamuk2012}
\bibinfo{author}{\bibfnamefont{B.}~\bibnamefont{Pamuk}},
  \bibinfo{author}{\bibfnamefont{J.~M.} \bibnamefont{Soler}},
  \bibinfo{author}{\bibfnamefont{R.}~\bibnamefont{Ram\'irez}},
  \bibinfo{author}{\bibfnamefont{C.~P.} \bibnamefont{Herrero}},
  \bibinfo{author}{\bibfnamefont{P.~W.} \bibnamefont{Stephens}},
  \bibinfo{author}{\bibfnamefont{P.~B.} \bibnamefont{Allen}}, \bibnamefont{and}
  \bibinfo{author}{\bibfnamefont{M.-V.} \bibnamefont{Fern\'andez-Serra}},
  \bibinfo{journal}{Phys. Rev. Lett.} \textbf{\bibinfo{volume}{108}},
  \bibinfo{pages}{193003} (\bibinfo{year}{2012}).

\bibitem[{\citenamefont{Murray and Galli}(2012)\citenamefont{E. D. Murray and Giulia Galli}}]{Galli2012}
\bibinfo{author}{\bibfnamefont{E. D.}~\bibnamefont{Murray}},
  \bibinfo{author}{\bibfnamefont{G.} \bibnamefont{Galli}},
  \bibinfo{journal}{Phys. Rev. Lett.} \textbf{\bibinfo{volume}{108}},
  \bibinfo{pages}{105502} (\bibinfo{year}{2012}).


\bibitem[{\citenamefont{Lin et~al.}(2011)\citenamefont{Lin, Morrone, Car, and
  Parrinello}}]{Lin2011}
\bibinfo{author}{\bibfnamefont{L.}~\bibnamefont{Lin}},
  \bibinfo{author}{\bibfnamefont{J.~A.}~\bibnamefont{Morrone}},
  \bibinfo{author}{\bibfnamefont{R.}~\bibnamefont{Car}}, \bibnamefont{and}
  \bibinfo{author}{\bibfnamefont{M.}~\bibnamefont{Parrinello}},
  \bibinfo{journal}{Phys. Rev. B} \textbf{\bibinfo{volume}{83}},
  \bibinfo{pages}{220302(R)} (\bibinfo{year}{2011}).


\bibitem[{\citenamefont{R\"{o}ttger et~al.}(1994)\citenamefont{R\"{o}ttger,
 Endriss, Ihringer,Doyle, and Kuhs}}]{Kuhs94}
\bibinfo{author}{\bibfnamefont{K.}~\bibnamefont{R\"{o}ttger}}, 
\bibinfo{author}{\bibfnamefont{A.}~\bibnamefont{Endriss}},
\bibinfo{author}{\bibfnamefont{J.}~\bibnamefont{Ihringer}},
\bibinfo{author}{\bibfnamefont{S.}~\bibnamefont{Doyle}}, \bibnamefont{and}
\bibinfo{author}{\bibfnamefont{W.~F.}~\bibnamefont{Kuhs}},
\bibinfo{journal}{Acta Cryst.}\textbf{\bibinfo{volume}{B50}},
\bibinfo{pages}{644} (\bibinfo{year}{1994}).

\bibitem[{\citenamefont{Giuliani et~al.}(2012)\citenamefont{Giuliani,
Ricci, Bruni and Mayers}}]{Mayers12}
\bibinfo{author}{\bibfnamefont{A.}~\bibnamefont{Giuliani}},
  \bibinfo{author}{\bibfnamefont{M.~A.} \bibnamefont{Ricci}},
  \bibinfo{author}{\bibfnamefont{F.}~\bibnamefont{Bruni}}, \bibnamefont{and}
  \bibinfo{author}{\bibfnamefont{J.}~\bibnamefont{Mayers}},
  \bibinfo{journal}{Phys. Rev. B} \textbf{\bibinfo{volume}{86}},
  \bibinfo{pages}{104308} (\bibinfo{year}{2012}).





\bibitem[{\citenamefont{Lin et~al.}(2010)\citenamefont{Lin, Morrone, Car, and
  Parrinello}}]{car10}
\bibinfo{author}{\bibfnamefont{L.}~\bibnamefont{Lin}},
  \bibinfo{author}{\bibfnamefont{J.~A.} \bibnamefont{Morrone}},
  \bibinfo{author}{\bibfnamefont{R.}~\bibnamefont{Car}}, \bibnamefont{and}
  \bibinfo{author}{\bibfnamefont{M.}~\bibnamefont{Parrinello}},
  \bibinfo{journal}{Phys. Rev. Lett.} \textbf{\bibinfo{volume}{105}},
  \bibinfo{pages}{110602} (\bibinfo{year}{2010}).



\bibitem[{\citenamefont{Kong et~al.}(2012)\citenamefont{Kong, Wu and Car}}]{Kong12}
\bibinfo{author}{\bibfnamefont{L.}~\bibnamefont{Kong}},
  \bibinfo{author}{\bibfnamefont{X.} \bibnamefont{Wu}},\bibnamefont{and}
  \bibinfo{author}{\bibfnamefont{R.}~\bibnamefont{Car}}, 
  \bibinfo{journal}{Phys. Rev. B.} \textbf{\bibinfo{volume}{86}},
  \bibinfo{pages}{134203} (\bibinfo{year}{2012}).

\bibitem[{\citenamefont{Ljungberg et~al.}(2010)\citenamefont{Ljungberg, Nilsson, and Pettersson}}]{Pettersson10}
\bibinfo{author}{\bibfnamefont{M.P.}~\bibnamefont{Ljungberg}},
  \bibinfo{author}{\bibfnamefont{A.} \bibnamefont{Nilsson}},\bibnamefont{and}
  \bibinfo{author}{\bibfnamefont{L.G.M.}~\bibnamefont{Pettersson}}, 
  \bibinfo{journal}{Phys. Rev. B.} \textbf{\bibinfo{volume}{82}},
  \bibinfo{pages}{245115} (\bibinfo{year}{2010}).


\bibitem[{\citenamefont{Fern\'andez-Serra and Artacho}(2006)}]{mvfsprl06}
\bibinfo{author}{\bibfnamefont{M.~V.} \bibnamefont{Fern\'andez-Serra}}
  \bibnamefont{and} \bibinfo{author}{\bibfnamefont{E.}~\bibnamefont{Artacho}},
  \bibinfo{journal}{Phys. Rev. Lett.} \textbf{\bibinfo{volume}{96}},
  \bibinfo{pages}{016404} (\bibinfo{year}{2006}).

\bibitem[{\citenamefont{Libowitzky}(1999)}]{Libowitzky99}
\bibinfo{author}{\bibfnamefont{E.}~\bibnamefont{Libowitzky}},
  \bibinfo{journal}{Monatshefte f\"{u}r Chemie}
  \textbf{\bibinfo{volume}{130}}, \bibinfo{pages}{1047} (\bibinfo{year}{1999}).

\bibitem[{\citenamefont{Ceriotti
  et~al.}(2009{\natexlab{a}})\citenamefont{Ceriotti, Bussi, and
  Parrinello}}]{ceriotti1}
\bibinfo{author}{\bibfnamefont{M.}~\bibnamefont{Ceriotti}},
  \bibinfo{author}{\bibfnamefont{G.}~\bibnamefont{Bussi}}, \bibnamefont{and}
  \bibinfo{author}{\bibfnamefont{M.}~\bibnamefont{Parrinello}},
  \bibinfo{journal}{Phys. Rev. Lett.} \textbf{\bibinfo{volume}{102}},
  \bibinfo{pages}{020601} (\bibinfo{year}{2009}{\natexlab{a}}).

\bibitem[{\citenamefont{Ceriotti
  et~al.}(2009{\natexlab{b}})\citenamefont{Ceriotti, Bussi, and
  Parrinello}}]{ceriotti2}
\bibinfo{author}{\bibfnamefont{M.}~\bibnamefont{Ceriotti}},
  \bibinfo{author}{\bibfnamefont{G.}~\bibnamefont{Bussi}}, \bibnamefont{and}
  \bibinfo{author}{\bibfnamefont{M.}~\bibnamefont{Parrinello}},
  \bibinfo{journal}{Phys. Rev. Lett.} \textbf{\bibinfo{volume}{103}},
  \bibinfo{pages}{030603} (\bibinfo{year}{2009}{\natexlab{b}}).

\bibitem[{\citenamefont{Ceriotti
  et~al.}(2010{\natexlab{a}})\citenamefont{Ceriotti, Bussi, and
  Parrinello}}]{ceriotti3}
\bibinfo{author}{\bibfnamefont{M.}~\bibnamefont{Ceriotti}},
  \bibinfo{author}{\bibfnamefont{G.}~\bibnamefont{Bussi}}, \bibnamefont{and}
  \bibinfo{author}{\bibfnamefont{M.}~\bibnamefont{Parrinello}},
  \bibinfo{journal}{J. Chem. Theory Comput.} \textbf{\bibinfo{volume}{6}},
  \bibinfo{pages}{1170} (\bibinfo{year}{2010}{\natexlab{a}}).

\bibitem[{\citenamefont{Ceriotti and Parrinello}(2010)}]{ceriotti4}
\bibinfo{author}{\bibfnamefont{M.}~\bibnamefont{Ceriotti}} \bibnamefont{and}
  \bibinfo{author}{\bibfnamefont{M.}~\bibnamefont{Parrinello}},
  \bibinfo{journal}{Proc. Comp. Sci.} \textbf{\bibinfo{volume}{1}},
  \bibinfo{pages}{1601} (\bibinfo{year}{2010}).

\bibitem[{\citenamefont{Ceriotti
  et~al.}(2010{\natexlab{b}})\citenamefont{Ceriotti, Parrinello, Markland, and
  Manolopoulos}}]{ceriotti5}
\bibinfo{author}{\bibfnamefont{M.}~\bibnamefont{Ceriotti}},
  \bibinfo{author}{\bibfnamefont{M.}~\bibnamefont{Parrinello}},
  \bibinfo{author}{\bibfnamefont{T.~E.} \bibnamefont{Markland}},
  \bibnamefont{and} \bibinfo{author}{\bibfnamefont{D.~E.}
  \bibnamefont{Manolopoulos}}, \bibinfo{journal}{J. Chem. Phys.}
  \textbf{\bibinfo{volume}{133}}, \bibinfo{pages}{124104}
  (\bibinfo{year}{2010}{\natexlab{b}}).

\bibitem[{\citenamefont{Morrone et~al.}(2011)\citenamefont{Morrone, Markland,
  Ceriotti, and Berne}}]{ceriotti6}
\bibinfo{author}{\bibfnamefont{J.~A.} \bibnamefont{Morrone}},
  \bibinfo{author}{\bibfnamefont{T.~E.} \bibnamefont{Markland}},
  \bibinfo{author}{\bibfnamefont{M.}~\bibnamefont{Ceriotti}}, \bibnamefont{and}
  \bibinfo{author}{\bibfnamefont{B.~J.} \bibnamefont{Berne}},
  \bibinfo{journal}{J. Chem. Phys.} \textbf{\bibinfo{volume}{134}},
  \bibinfo{pages}{084104} (\bibinfo{year}{2011}).

\bibitem[{\citenamefont{Raghunathan et~al.}(2011)\citenamefont{Raghunathan,
  Greaney, and Grossman}}]{Phonostat-Grossman}
\bibinfo{author}{\bibfnamefont{R.}~\bibnamefont{Raghunathan}},
  \bibinfo{author}{\bibfnamefont{P.~A.} \bibnamefont{Greaney}},
  \bibnamefont{and} \bibinfo{author}{\bibfnamefont{J.~C.}
  \bibnamefont{Grossman}}, \bibinfo{journal}{J. Chem. Phys.}
  \textbf{\bibinfo{volume}{134}}, \bibinfo{eid}{214117}
  (pages~\bibinfo{numpages}{9}) (\bibinfo{year}{2011}).

\bibitem[{\citenamefont{Ford et~al.}(1965)\citenamefont{Ford, Kac, and
  Mazur}}]{fordmazur}
\bibinfo{author}{\bibfnamefont{G.~W.} \bibnamefont{Ford}},
  \bibinfo{author}{\bibfnamefont{M.}~\bibnamefont{Kac}}, \bibnamefont{and}
  \bibinfo{author}{\bibfnamefont{P.}~\bibnamefont{Mazur}},
  \bibinfo{journal}{Journal of Mathematical Physics}
  \textbf{\bibinfo{volume}{6}}, \bibinfo{pages}{504} (\bibinfo{year}{1965}).

\bibitem[{\citenamefont{Zwanzig}(1973)}]{zwanzig1973}
\bibinfo{author}{\bibfnamefont{R.}~\bibnamefont{Zwanzig}},
  \bibinfo{journal}{Journal of Statistical Physics}
  \textbf{\bibinfo{volume}{9}}, \bibinfo{pages}{215} (\bibinfo{year}{1973}).

\bibitem [{note1()}]{note1}%
  \BibitemOpen
   {}\bibinfo {note} {Note that the prescription to write Eq.~(\ref{eq:1dgle-2}) as higher dimensional Markovian process is a separate procedure and the auxiliary momenta $s$ do not have any connection to the bath degrees of freedom (in this analysis). The auxiliary momenta $s$ are introduced only for implementation purposes.}\BibitemShut
  {Stop}%


\bibitem[{\citenamefont{Bussi and Parrinello}(2007)}]{bussi}
\bibinfo{author}{\bibfnamefont{G.}~\bibnamefont{Bussi}} \bibnamefont{and}
  \bibinfo{author}{\bibfnamefont{M.}~\bibnamefont{Parrinello}},
  \bibinfo{journal}{Phys. Rev. E} \textbf{\bibinfo{volume}{75}},
  \bibinfo{pages}{056707} (\bibinfo{year}{2007}).

\bibitem[{\citenamefont{Tuckerman et~al.}(1992)\citenamefont{Tuckerman, Berne,
  and Martyna}}]{tuckerman1990}
\bibinfo{author}{\bibfnamefont{M.}~\bibnamefont{Tuckerman}},
  \bibinfo{author}{\bibfnamefont{B.~J.} \bibnamefont{Berne}}, \bibnamefont{and}
  \bibinfo{author}{\bibfnamefont{G.~J.} \bibnamefont{Martyna}},
  \bibinfo{journal}{J. Chem. Phys.}
  \textbf{\bibinfo{volume}{97}}, \bibinfo{pages}{1990} (\bibinfo{year}{1992}).

\bibitem[{\citenamefont{R.~Ram\'{\i}rez et~al.}(2012)\citenamefont{Ram\'irez,
 Neuerburg, Fern\'andez-Serra, and Herrero}}]{Ramirez12}
 \bibinfo{author}{\bibfnamefont{R.}~\bibnamefont{Ram\'irez}},
 \bibinfo{author}{\bibfnamefont{N.}~\bibnamefont{Neuerburg}},
  \bibinfo{author}{\bibfnamefont{M.-V.} \bibnamefont{Fern\'andez-Serra}}, \bibnamefont{and}
  \bibinfo{author}{\bibfnamefont{C.}~\bibnamefont{Herrero}},
  \bibinfo{journal}{J. Chem. Phys.}
  \textbf{\bibinfo{volume}{137}}, \bibinfo{eid}{044502}
  (pages~\bibinfo{numpages}{11}) (\bibinfo{year}{2012}).

\bibitem [{note2()}]{note2}%
  \BibitemOpen
   {}\bibinfo {note} {http://gle4md.berlios.de}\BibitemShut
  {Stop}%

\bibitem[{\citenamefont{Hardy et~al.}(1998)}]{Hardy98}
\bibinfo{author}{\bibfnamefont{R.J.}~\bibnamefont{Hardy}},
\bibinfo{author}{\bibfnamefont{N.J.}~\bibnamefont{Lacks}} \bibnamefont{and}
  \bibinfo{author}{\bibfnamefont{R.C.}~\bibnamefont{Shukla}},
  \bibinfo{journal}{Phys. Rev. B} \textbf{\bibinfo{volume}{57}},
  \bibinfo{pages}{833} (\bibinfo{year}{1998}).

\bibitem[{\citenamefont{Habershon et~al.}(2008)\citenamefont{Habershon,
  Fanourgakis, and Manolopoulos}}]{pimdspectrum}
\bibinfo{author}{\bibfnamefont{S.}~\bibnamefont{Habershon}},
  \bibinfo{author}{\bibfnamefont{G.~S.} \bibnamefont{Fanourgakis}},
  \bibnamefont{and} \bibinfo{author}{\bibfnamefont{D.~E.}
  \bibnamefont{Manolopoulos}}, \bibinfo{journal}{J. Chem. Phys.} \textbf{\bibinfo{volume}{129}}, \bibinfo{eid}{074501}
  (pages~\bibinfo{numpages}{8}) (\bibinfo{year}{2008}).

\bibitem[{\citenamefont{Habershon and Manolopoulos}(2009)}]{zpleakage}
\bibinfo{author}{\bibfnamefont{S.}~\bibnamefont{Habershon}} \bibnamefont{and}
  \bibinfo{author}{\bibfnamefont{D.~E.} \bibnamefont{Manolopoulos}},
  \bibinfo{journal}{J. Chem. Phys.}
  \textbf{\bibinfo{volume}{131}}, \bibinfo{eid}{244518}
  (pages~\bibinfo{numpages}{11}) (\bibinfo{year}{2009}).


\bibitem[{\citenamefont{Poma and Delle Site}(2010)}]{Poma10}
\bibinfo{author}{\bibfnamefont{A.B.}\bibnamefont{Poma}} \bibnamefont{and}
  \bibinfo{author}{\bibfnamefont{L.}\bibnamefont{Delle Site}},
  \bibinfo{journal}{Phys. Rev. Lett}
  \textbf{\bibinfo{volume}{104}}, \bibinfo{eid}{250201}
  (pages~\bibinfo{numpages}{4}) (\bibinfo{year}{2010}).

\end{thebibliography}
\end{document}